\documentclass{article}

\usepackage{microtype}
\usepackage{graphicx}
\usepackage{subfigure}
\usepackage{booktabs} % for professional tables

\usepackage{hyperref}

\usepackage[accepted]{icml2025}

\usepackage{amsmath}
\usepackage{amssymb}
\usepackage{mathtools}
\usepackage{amsthm}

\usepackage[capitalize,noabbrev]{cleveref}

\usepackage{url}
\usepackage{acronym}
\graphicspath{{figures/}}
\usepackage{balance}
\usepackage{xspace}
\usepackage{setspace}
\usepackage{tabularx,colortbl,multirow,multicol,array,makecell,tabularray}
\usepackage{overpic,wrapfig}
\usepackage[misc]{ifsym}
\usepackage{pifont}
\usepackage{bbding}
\usepackage{arydshln}

\acrodef{llm}[LLM]{Large Language Model}
\acrodef{mllm}[MLLM]{Multimodal Large Language Model}
\acrodef{hri}[HRI]{Human-Robot Interaction}

\makeatletter
\DeclareRobustCommand\onedot{\futurelet\@let@token\@onedot}
\def\@onedot{\ifx\@let@token.\else.\null\fi\xspace}
\def\eg{e.g\onedot} 

\def\ie{i.e\onedot}

\makeatother

\theoremstyle{plain}

\theoremstyle{definition}

\theoremstyle{remark}

\usepackage[textsize=tiny]{todonotes}

\icmltitlerunning{EmpathyAgent: Can Embodied Agents Conduct Empathetic Actions?}

\begin{document}

\twocolumn[
\icmltitle{EmpathyAgent: Can Embodied Agents Conduct Empathetic Actions?}



\icmlsetsymbol{equal}{*}

% \begin{icmlauthorlist}
% \icmlauthor{Firstname1 Lastname1}{equal,yyy}
% \icmlauthor{Firstname2 Lastname2}{equal,yyy,comp}
% \icmlauthor{Firstname3 Lastname3}{comp}
% \end{icmlauthorlist}

\icmlsetsymbol{equal}{*}
\begin{icmlauthorlist}
\icmlauthor{Xinyan Chen$^{*}$}{PKU}\textbf{, }
\icmlauthor{Jiaxin Ge$^{*}$}{PKU}\textbf{, }
\icmlauthor{Hongming Dai$^{*}$}{PKU}\textbf{, }
\icmlauthor{Qiang Zhou}{PITT}\textbf{, }
\icmlauthor{Qiuxuan Feng$^{}$}{PKU}\textbf{, }
\icmlauthor{Jingtong Hu}{PITT}\textbf{, }
\icmlauthor{Yizhou Wang}{PKU}\textbf{, }
\icmlauthor{Jiaming Liu}{PKU}\textbf{, }
\icmlauthor{Shanghang Zhang {\Envelope}}{PKU}\\
\vspace{0.2cm}
\begin{tabular}{c}
\textsuperscript{1} State Key Laboratory of Multimedia Information Processing, School of Computer Science, Peking University\\
\textsuperscript{2} University of Pittsburgh\\
\vspace{0.5em}
\texttt{shanghang@pku.edu.cn}\\
\vspace{0.1cm}
$^{*}$ Equal contribution\quad
% $^{\ddagger}$ Internship at PKU\quad
\Envelope ~Corresponding author
\end{tabular}
\end{icmlauthorlist}

\vskip 0.3in
]




\begin{figure*}[ht]
    \centering
    \includegraphics[width=0.9\linewidth]{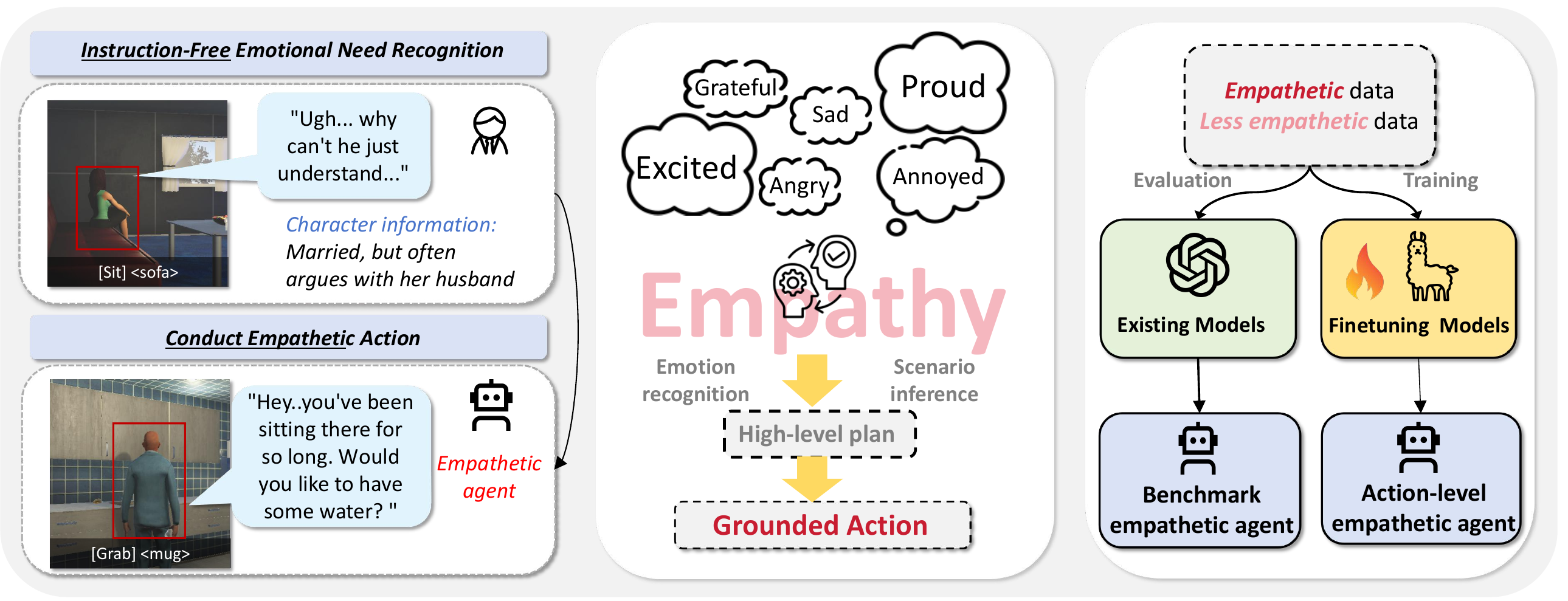}
    % \vskip 0.1in
    \caption{\textbf{We propose EmpathyAgent, the first benchmark to evaluate and enhance the empathetic actions of embodied agents.} In a simulated environment, the embodied agent is tasked to observe a scenario and then perform responsive empathetic actions. In this example, the agent first observes that there is a person sitting on the sofa and sighing. Considering the background information, the agent realizes the person is sad, and then conducts the action of bringing some water to the person.
    Meanwhile, EmpathyAgent can also be used to train embodied agents and boost empathetic behaviors.}
    % \vskip 0.1in
    \label{fig:overview}
\end{figure*}

\begin{abstract}
Empathy is fundamental to human interactions, yet it remains unclear whether embodied agents can provide human-like empathetic support. 
Existing works have studied agents' tasks solving and social interactions abilities, but whether agents can understand empathetic needs and conduct empathetic behaviors remains overlooked.
To address this, we introduce \textbf{EmpathyAgent}, the first benchmark to evaluate and enhance agents' empathetic actions across diverse scenarios. 
\textbf{EmpathyAgent} contains 10,000 multimodal samples with corresponding empathetic task plans and three different challenges. 
To systematically evaluate the agents' empathetic actions, we propose an empathy-specific evaluation suite that evaluates the agents' empathy process.
We benchmark current models and found that exhibiting empathetic actions remains a significant challenge.
Meanwhile, we train Llama3-8B using \textbf{EmpathyAgent} and find it can potentially enhance empathetic behavior.
By establishing a standard benchmark for evaluating empathetic actions, we hope to advance research in empathetic embodied agents.
Our code and data are publicly available at \url{https://github.com/xinyan-cxy/EmpathyAgent}.
\end{abstract}

\section{Introduction}
% Empathy Important
% Empathy is important in human society (big picture) 
% Empathy is \cite{empathydefinition}
% Strong paragraph of why empathy is important
\begin{quote}
    ``No quality of human nature is more remarkable, both in itself and in its consequences, than that propensity we have to sympathize with others, and to receive by communication their inclinations and sentiments, however different from, or even contrary to our own.'' 
    \begin{flushright}
    --- David Hume \cite{hume2000treatise}. 
    \end{flushright}
\end{quote}

%jiaxin
% Imagine you are terribly sick in a foreign country, you call a cab and go to a hospital, alone, feeling helpless and scared. While you were waiting anxiously outside, someone noticed you, recognized your pain, then came up to you and softly asked you if you needed a hug. You don't know this person at all, but such a simple action made you feel so much better...
% % give an intuitve example - ben
% Empathy is a fundamental instinct in human nature. Every person has a need in nature to see the happiness of others \cite{smith2010theory}. 
% Receiving empathetic support from others enables us to feel understood, valued, and accepted. 
% As AI increasingly integrates into daily life and impacts human society \cite{vicentini2021collaborative, garcia2007evolution, breazeal2016social}, a natural question emerges: Can such support come from an AI?

Imagine being terribly sick in a foreign country. You call a cab and go to the hospital alone, feeling helpless and scared. While waiting anxiously outside, someone notices you, recognizes your pain, and comes up to you, softly asking if you need a hug. You don’t know this person at all, but such a simple action makes you feel so much better...
% give an intuitive example - ben
This highlights how empathy is a fundamental instinct in human nature. Every person has a need in nature to see the happiness of others \cite{smith2010theory}. 
Receiving empathetic support from others enables us to feel understood, valued, and accepted. 
Recently, as embodied agents increasingly integrate into daily life \cite{brohan2022rt, huang2023voxposer, li2023manipllm} and become reliable assistant agents to people \cite{vicentini2021collaborative, breazeal2016social}, a natural question emerges: Can such support come from intelligent embodied agents?

Scientific-wise, studying to what extent embodied agents can behave empathetically helps us analyze how far these current models are from human intelligence.
Recent studies show that although these models are still far from being authentically conscious \cite{chalmers2023could}, they can exhibit certain theory of mind abilities \cite{strachan2024testing}. 
By studying how much these models can exhibit empathetic behaviors, we can understand how far these models are from human-level emotional intelligence.
% Applicational-wise
Application-wise, pushing intelligent agents to exhibit empathy enables them to better meet human needs and provide empathetic support \cite{leite2013influence, paiva2017empathy}.
Recent studies show that agents can make people ``feel heard,'' suggesting they have the potential to offer emotional and empathetic support to humans \cite{doi:10.1073/pnas.2319112121}. 
% Enabling current models to recognize emotional needs and conduct empathetic responses allows us to build models that meet humans' empathetic needs.
% Current Limitations
However, this field has been largely underexplored. There are no existing benchmarks that can systematically evaluate the ability of embodied agents to perform empathetic actions.
Existing benchmarks focus mainly on the success rate in completing a given task \cite{puig2020watch, shridhar2020alfred} and neglect the aspect of empathy.

%jiaxin
% In this paper, we present ``EmpathyAgent'', the first benchmark with 10,000 data points designed to evaluate and enhance the model's ability to conduct empathetic actions. This benchmark is built upon the VirtualHome simulator \cite{puig2018virtualhome}. This is a simulated home environment where the agent can perform many different actions to interact with it, such as picking up objects, switching on/off appliances, or opening appliances.
% We design various scenarios that contain a person in need of emotional support and let the agent conduct actions in response. Echoing core components of how humans perform empathetic actions \cite{preston2002empathy}, we design our EmpathyAgent based on the following principles:
% \textbf{First}, the agent needs to perceive the empathic cues (expression or situation) from the human.
% \textbf{Second}, the agent needs to engage in an internal affective or cognitive process to understand the scenario, such as what is the person's feeling and what may have happened to cause the person to behave this way.
% \textbf{Third}, the agent needs to convert such process to its internal outcomes, such as whether it should mirror the person's emotions and how to take the person's characteristics into account.
% \textbf{And finally}, the agent needs to plan and successfully execute the grounded actions as the empathetic response. Based on these steps, our scenario contains the background of the person; the person's action in the form of a video; and also the person's language; An example of our scenario can be found in \Cref{fig:pipeline}.

In this paper, we propose ``EmpathyAgent,'' the first benchmark consisting of 10k samples, with 1k testing samples and 9k training samples, designed to evaluate and enhance embodied agents' ability to perform empathetic actions. We demonstrate the overview of EmpathyAgent in Figure~\ref{fig:overview}.

\begin{figure*}[t]
    \centering
    \includegraphics[width=0.8\linewidth]{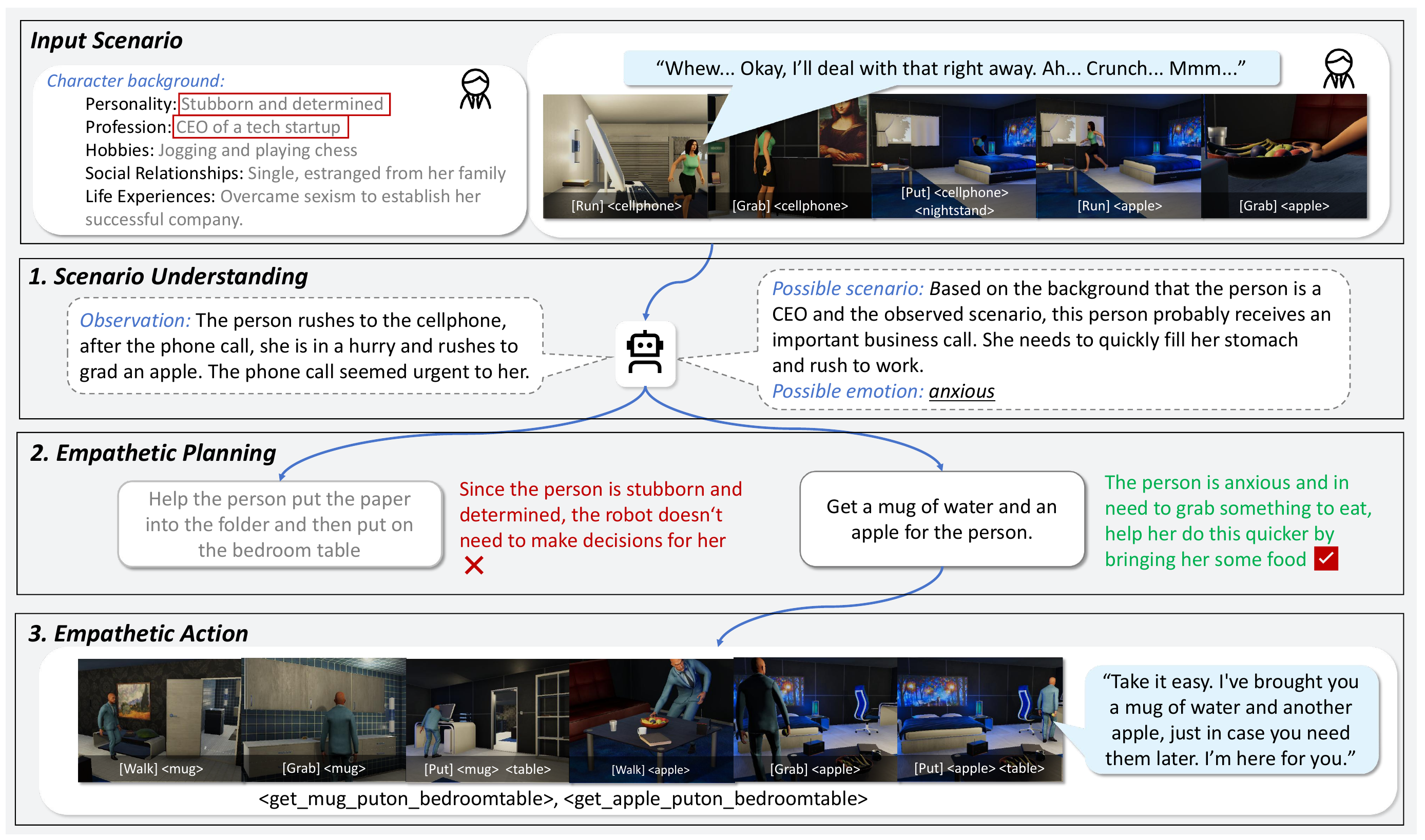}
    % \vskip 0.1in
    \caption{\textbf{An example of EmpathyAgent.} The agent is provided with an input scenario, which contains a character with a personal background; A video of the character taking a sequence of actions (\eg, rushing to get the phone and then get the apples); A language cue from the character (\eg, saying something while performing the actions).
    To perform empathetic actions after observing this scenario, the agent needs to conduct three steps: \textbf{(1) Scenario Understanding}: Based on the scenario, the agent goes through a cognitive and affective process to determine the emotional state of the person and the possible causes that lead to this state. \textbf{(2) Empathetic Planning}: Based on the understanding from the internal empathy process, the agent comes up with possible plans of what actions to conduct under this scenario. The agent should reason about which plan meets the empathetic needs of the person based on the person's personal background. \textbf{(3) Empathetic Actions}: Finally, based on the high-level plan, the agent outputs a series of grounded and executable empathetic actions and performs them in the environment.}
    % \vskip 0.1in
    \label{fig:pipeline}
\end{figure*}

EmpathyAgent is built upon the VirtualHome simulator \cite{puig2018virtualhome}, a simulated home environment where the agent can perform a wide range of actions, such as picking up objects, switching appliances on/off, or opening appliances.
We design various scenarios that involve a person in need of emotional support and let the embodied agent generate empathy-driven task planning in response.
We design EmpathyAgent based on the principles of how humans perform empathetic actions \cite{preston2002empathy} and the design principles of social robots \cite{park2022empathy}, which contains multiple steps of the empathy process for agents. Specifically, EmpathyAgent contains three challenges: 
\textbf{First}, the embodied agent needs to perceive empathetic cues (i.e., expressions or situations) from the human, The embodied agent engages in an internal affective or cognitive process to understand the scenario, such as determining the person's feelings and what might have caused their behavior.
\textbf{Second}, the agent converts such process to its internal outcomes, such as whether it should mirror the person's emotions and how to take the person's characteristics into account.
\textbf{Finally}, the embodied agent plans and executes a sequence of empathy-driven actions as its response. Based on these steps, our scenario contains the background of the person, the person's actions in the form of a video, and also the person's language. An example of our scenario can be found in \Cref{fig:pipeline}. 
% Experiment
After seeing the scenario, the agent takes a series of actions in response. For each data sample, we present two action sequences and use human feedback to label one as ``empathetic'' and the other as ``less empathetic''.

To systematically test agents' empathetic behavior, we design three challenges where the agent can access different amount of information.
We then develop a set of metrics to systematically evaluate a model's empathy capacity. Our metrics suite contains both (i) reference-based metrics, which measure the similarity between the agent's response and the reference response, and (ii) reference-free metrics, which evaluate the alignment of the agent's response with psychological principles of empathy.

We benchmark on various LLMs and VLMs such as GPT-4 \cite{openai2023gpt4} and Llama 3\cite{touvron2023llama}, and find that this benchmark remains challenging for current models.
To further demonstrate the practicality of EmpathyAgent, we finetune Llama3-8B and test its performance.
We find that training on our benchmark improves the model's performance in exhibiting empathetic behaviors. This suggests that EmpathyAgent can not only significantly enhance the embodied agent's ability to generate empathetic responses but also promote future research on building real-world empathetic embodied agents.
Our contributions are summarized as follows:
\vspace{-5pt}
\begin{itemize}
    \item We introduce \textbf{EmpathyAgent}, the first benchmark with 10k samples for evaluating and enhancing the empathetic actions of embodied agents. EmpathyAgent is designed to mimic the human empathy process and can be scaled up automatically. It makes a first attempt to advance the study of building embodied agents that provide empathetic support to humans.
    \item We develop a systematic evaluation framework based on the human empathy process, with both reference-based metrics and reference-free metrics. 
    \item We benchmark on the current LLM and VLM models. Additionally, we train Llama3-8B on EmpathyAgent and find it performs similarly with GPT-4-Turbo, demonstrating that \textbf{EmpathyAgent} can be used in enhancing the empathetic behaviors of embodied agents.
\end{itemize}

\section{Related Work}
\textbf{Human-Robot Interaction}
The study of \ac{hri} has a long history \cite{goodrich2008human}. Prior works have built simulated lab environments to conduct such studies \cite{7450630}, which greatly limits diversity and generalization. Recently, Watch-And-Help \cite{puig2020watch} develops a simulated home environment with various objects, and the agent can perform actions to help the person complete a task. Communicative Watch-And-Help \cite{zhang2023building} adds a communicative channel where the agents can interact through language to better perform cooperative actions. However, these works are focused on better task completion (\ie, putting an apple on the plate) but fail to consider how empathy affects human-robot interactions. We study how embodied agents can recognize empathetic needs and perform empathetic actions even when no explicit task instruction is given.

\textbf{LLMs and MLLMs in Embodied Agents} 
Recently, \acp{llm} and \acp{mllm} \cite{li2023blip, liu2023visual, alayrac2022flamingo} have been used for robotics control and planning.  SayCan \cite{ahn2022can} uses \ac{llm} to interpret high-level task instructions and then forms detailed low-level language instructions that can be directly mapped to the embodied agent's low-level actions to complete the task. PaLM-E \cite{driess2023palm} uses a multimodal language model for embodied reasoning. \cite{wang2024find} uses \acp{llm} to do visual navigation to find objects on the user's demand. However, these works are more focused on successfully performing certain actions for a given task (\eg, finding the water or picking up an object). They neglect the aspect of studying social interactions between agents.

% \section{Overview of EmpathyAgent}
% ? or make this a subsection of the {Dataset Generation Method} one, but rename {Dataset Generation Method} --> something like {designing EmpathyAgents?}. like you need to give some concrete details for i). what are the tasks ii). what are the inputs/outputs of the models iii). why each task is important / where they come from. --- by ben
% \section{Method}
\section{EmpathyAgent}
In this section, we first provide the task formulation of EmpathyAgent (Sec. \ref{sec:task_formulation}). Then, we introduce our pipeline for creating the benchmark (Sec. \ref{sec:benchmark_generation}). 
% Next, we introduce our evaluation metrics (Sec. \ref{sec:eval_metrics}). Finally, we introduce our method of using EmpathyAgent benchmark to train empathetic agents (Sec. \ref{sec:train_method}).

\subsection{Task Formulation}
\label{sec:task_formulation}
Our task is defined as follows: Given a scenario, the agent needs to perform grounded actions that are empathetically responsive to the human, as shown in Figure 2.
% \subsubsection{Input Scenario}
% \vspace{-5pt}
Similar to what humans can observe in real-world interactions, the input consists of
three parts: The basic background of the person in the scenario, the video of the person performing actions in the scenario, and what the person says in the scenario.
% \subsubsection{Three Output Challenges}
% \vspace{-5pt}
\label{sec:three_challenges}
For the output part, we design three
challenges for the agent based on the three steps that the agent needs to take in order to successfully demonstrate empathetic behaviors. 
\vspace{-5pt}
\paragraph{Scenario Understanding} 
Understanding the scenario is the first and most basic step towards empathetic behavior. 
The agent perceives the scenario, understands the content of the scene, and reasons about the underlying facts behind the scenario.
% such as what may have caused the person to perform these actions, and what is the person’s underlying emotions. 
To formulate this as a challenge, the agent receives the character’s input actions in the scenario and the character’s background information. 
The agent is tasked to output a scenario description of these components based on its understanding of the scenario. This tests the agent's understanding of the scenario.
% The agent should output its understanding of the scenario. This includes recognizing the person’s emotions and identifying the possible causes. 
\vspace{-5pt}
\paragraph{Empathetic Planning} Then, the agent should formulate a high-level plan of what to do after comprehending the scenario. For example, after noticing the person hasn’t eaten anything because of being too upset, the agent may come up with a plan like “Find the person some of his favorite food, then comfort him.” To formulate this process as a challenge, the agent receives the character’s input actions and the character’s background information. Then the agent is tasked to output such empathetic planning. 
% the agent should output a high-level plan of what it should do in the scenario. This includes understanding and reasoning about what possible responses are empathetic and responsive.
\vspace{-5pt}
\paragraph{Empathetic Actions} Finally, the agent should translate the high-level plan into grounded, low-level actions supported in the VirtualHome environment. For example, the high-level plan “Find the person some of his favorite food” might be grounded to “go to table”, “take chocolate bar”, “go to bedroom”, “put a chocolate bar on bedroom table”.
To formulate this as a challenge, the agent receives the character’s input actions, the character’s background information, and instructions for the low-level actions executable in the simulated environment. Then, the agent generates empathetic actions in a specific format.
% The agent should output grounded, executable actions in the simulated environment. This includes taking valid actions in the environment, such as walking somewhere, picking up an object or saying something.

\subsection{Benchmark Creation}
\label{sec:benchmark_generation}
\begin{figure*}[t]
    \centering
    \includegraphics[width=0.8\linewidth]{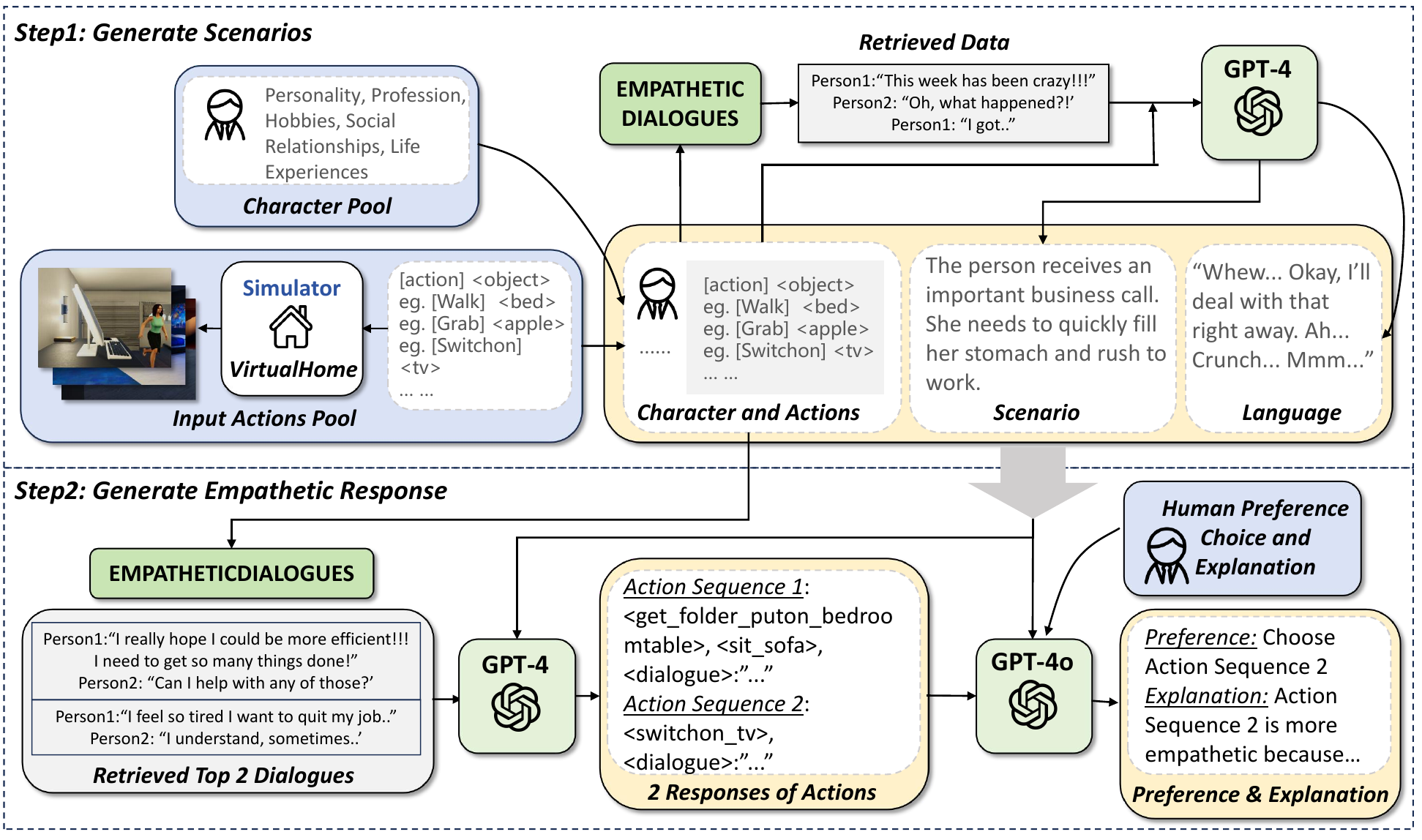}
    % \vskip 0.1in
    \caption{\textbf{Benchmark creation pipeline.} 
    \textbf{Step1}, we generate diverse scenarios. To do this, we sample a character and the character's input action. We use them to retrieve data from EmpatheticDialogues and use them together to generate a scenario description and the person's dialogue. The retrieval step ensures the generated scenario's diversity. \textbf{Step2}, we generate an empathetic response for each scenario. To do this, we use the scenario to retrieve the top two data points from the EmpatheticDialogues and use each of them as a source to generate a corresponding empathetic response. We then let the model choose the more empathetic response by using human-annotated examples and explanations as in-context examples. In this way, we construct a paired empathetic response where one is labeled more empathetic and the other is labeled less.}
    % \vskip 0.1in
    \label{fig:method_pipeline}
\end{figure*}
In this part, we describe the EmpathyAgent creation pipeline. This contains the input generation part and the output generation part. An overview of the pipeline is presented in \Cref{fig:method_pipeline}.
\vspace{-5pt}
\paragraph{Scenario Generation}
First, we generate diverse scenarios that contain a person in need of empathetic support. This process involves three steps:
\textbf{(i) Character Pool Generation.} We begin by creating a character pool containing diverse characters. Each character has a set of attributes, including Personality, Profession, Hobbies, Social Relationships, and Life Experiences. This diversity ensures that the scenarios cover a wide range of human behaviors and contexts.
\textbf{(ii) Input Actions Pool Definition.} We define an Input Actions Pool that contains various valid action sequences that an agent may take in the VirtualHome environment, such as pacing back and forth.
\textbf{(iii) Scenario Creation.} From the character pool and the input actions pool, we sample a pair that contains a character and a series of actions this character conducts. 
We then use this pair as input information and use GPT-4-turbo \cite{openai2023gpt4} to generate a scenario and the character's dialogue when conducting these actions. 
To ensure the generated scenario's diversity, following previous work \cite{zhou2023sotopia}, we use retrieval-augmented generation \cite{lewis2020retrieval} to retrieve the most relevant data point from an external dataset. In this case, we use EmpatheticDialogues \cite{rashkin-etal-2019-towards}, a dataset containing human dialogues that show empathy. We use the retrieved data as additional input information to generate the scenario. 
\vspace{-5pt}
\paragraph{Empathy Response Generation}
Second, we generate empathetic action sequences for each scenario and create labels for them. This involves two steps: \textbf{(i) Action Generation.} For each scenario, we first retrieve the two most relevant data points from EmpatheticDialogues and use them to separately generate two output actions. The legal action space is provided to the model so that the model can only choose possible actions.
% \textbf{(i) Action Generation} 
% We predefined a pool of 51 action sequences in the format like '<get\_folder\_puton\_bedroomtable>' and '<sit\_sofa>'. Those action sequences can be directly transformed into action lists that can be input into VirtualHome, like '[Grab] <folder>'.
% To generate action responses, we first use the entire input information (character background, actions, scenario description, language) to retrieve the top-2 most relevant data samples from EmpathyDialogue. Then, we use these two samples and the input information to let GPT-4 select from the action sequences pool and generate two empathetic output action responses that a robot may take under this scenario. This process enables the diversity of possible actions and also provides examples of empathetic interactions. 
% \textbf{(i) High-level Plan Generation} We also combined the entire input information and the two empathetic output actions to generate two high-level planning responses for each data point. This process ensures the alignment of action-level responses and high-level responses of our dataset.
\textbf{(ii) Action Selection.} Then, we label the preference between the two responses. To do this, we first construct some in-context examples labeled by human annotators. The annotators are asked to choose the more empathetic response based on the input information and then write an explanation of their choice. We next use these human-annotated examples and let GPT-4o select the more empathetic response and provide an explanation for the choice. We will provide more details of the benchmark in \Cref{sec:dataset_details}.

\section{Evaluation Metrics}
\label{sec:eval_metrics}
% In this part, we introduce the evaluation metrics that we designed and our design principles
In this section, we introduce our evaluation framework for EmapthyAgent, which consists of reference-based metrics aligned with the reference outputs of the three challenges, and reference-free metrics grounded in principles from psychology and Human-Robot Interaction (HRI).
\subsection{Reference-Based Metrics}
\vspace{-5pt}
Our reference-based framework is structured based on the three output challenges in Sec. \ref{sec:three_challenges}.\\
\textbf{Scenario Understanding} 
The scenario understanding process requires models to output their understanding of the scenario.
To evaluate this process, 
we compare the model's output scenario description with the ground-truth scenario description. We use the standard NLG metrics including Bleu (from Bleu-1 to Bleu-4) \cite{papineni-etal-2002-bleu}, ROUGE-L \cite{lin-2004-rouge}, CIDEr \cite{vedantam2015cider} and SPICE \cite{anderson2016spice}. We also use BERTScore \cite{zhang2020bertscore} which computes embeddings' similarity.\\
\textbf{Empathetic Planning}
In the empathetic planning process, models output a high-level plan of what they should do in the scenario. For evaluation, 
we compare the model's output plan with the ground-truth plan. We use the same NLG metrics as in Scenario Understanding.\\
\textbf{Empathetic Actions}
The empathetic actions process requires models to output grounded, executable actions in the simulated environment.
To evaluate this process, we use Overlap and TF-IDF scores between the model's actions and the ground-truth actions. 
Overlap computes the action overlapping rate between the output sequence and the ground-truth sequence. And TF-IDF computes the weighted similarity between the actions in the output sequence and the ground-truth sequence.
Following VirtualHome \cite{puig2018virtualhome}, we also use the LCS (Longest Common Subsequence) metric. 
LCS computes the longest common subsequence length between the output action sequence and the ground-truth action sequence. Additional details are provided in \Cref{sec:NLG_metric_details}.

\subsection{Reference-Free Metrics}
We design a set of reference-free metrics that draws on insights from psychology and HRI. Inspired by the RoPE scale \cite{charrier2019rope} metric that measures the perception of a robot's empathy from a second-person perspective in HRI, we design our reference-free metrics on eight dimensions to evaluate the three challenges of the agent's empathy response. We specify more details of how the dimensions in our evaluation framework correspond with the RoPE scale in \Cref{sec:Correspondence}.
\begin{itemize}
\setlength{\itemsep}{-2pt}
\item \textbf{Action and Dialogue Association} assesses the agent's ability to understand the underlying information of the character's actions and dialogues.
This metric is motivated by the cognitive process of empathy \cite{park2022empathy}.
\item \textbf{Individual Understanding} assesses whether the agent takes into account all details of a character's background information, and deducts the character's perspective based on it. This metric is motivated by the perspective-taking process \cite{park2022empathy}.
\item \textbf{Emotional Communication} evaluates (1) whether the emotion recognition is appropriate and (2) whether the agent expresses appropriate emotion. This metric is motivated by the Feature-Level Evaluation in \cite{yalccin2019evaluating}
\item \textbf{Emotion Regulation} evaluates whether the agent helps with the emotion. This metric is also based on \cite{yalccin2019evaluating}.
\item \textbf{Helpfulness} evaluates whether the agent effectively assists the character.
\item \textbf{Adaptability} evaluates the agent's flexibility and responsiveness in diverse scenarios. It evaluates whether the agent's interaction with the character is perceived as comfortable. This is an important aspect \cite{charrier2018empathy} in HRI;
\item \textbf{Coherence} evaluates the agent's consistency. This includes logical consistency such as whether the agent's understanding
of the scenario is consistent over time, and also action consistency such as whether the action matches the understanding.
\item \textbf{Legality} assesses whether the action sequence is legal and executable.
\end{itemize}
% It is applied to evaluate all three steps.\\
For each challenge, we apply distinct metric dimensions, as detailed in \Cref{tab:metric_usage}.
To evaluate different models using this set of metrics, we follow \cite{zhou2023sotopia} to use GPT-4-turbo to score between [1-10] on each dimension. The prompts we used for GPT-4-turbo to score on each dimension are presented in \Cref{sec:instructions}. To validate the reliability of GPT-based evaluation, we sample 50 data points and conduct a human evaluation across three challenges, where human annotators follow the same prompts as GPT-4-turbo to rate each dimension except for \textit{Legality}. We then compute the average intraclass correlation coefficient (ICC) between human ratings and GPT scores. The average ICC score is 74.76\% with a 95\% confidence interval, confirming the high reliability of our GPT-based evaluation.
% \subsection{Empathetic Agent Training Method}
\section{Training Method}
\label{sec:train_method}
We then leveraged the EmpathyAgent benchmark to train an empathetic agent and see whether it could output empathetic responses. We used the full training set and trained on Llama3-8B \cite{touvron2023llama} using two approaches: (1) Instruction finetuning and (2) Reinforcement Learning with Human Feedback (RLHF)  \cite{ouyang2022training}. For instruction finetuning, we used the response that was labeled as ``more empathetic'' as the ground truth and used the LoRA technique \cite{hu2021lora} to finetune the model. For RLHF, we first used our paired data to train a reward model, and then we used this reward model to train Llama3-8B using LoRA. 
% Specifically, we employed the Proximal Policy Optimization (PPO) algorithm \cite{schulman2017proximal} for policy optimization. In this setup, the reward model acts as a guiding signal to encourage the generation of more empathetic responses, with the reward at each time step \( r_t \) calculated as \( r_t = \text{RewardModel}(s_t, a_t) \), where \( s_t \) denotes the state (e.g., conversation context) and \( a_t \) the action (e.g., model’s response).

% \[
% L_{\text{PPO}}(\theta) = \mathbb{E}_t \left[ \min \left( r_t \cdot \hat{A}_t, \text{clip}(r_t, 1 - \epsilon, 1 + \epsilon) \cdot \hat{A}_t \right) \right]
% \]

% where \( \hat{A}_t \) is the advantage estimate at time step \( t \), and \( \epsilon \) is the clipping parameter that prevents large updates to the model's policy.
% By conducting these two experiments, we explore whether and to what extent this benchmark can be used to leverage empathetic responses in current agents. 

\begin{table}[t]
\small
\centering
\caption{Metric usage for different challenges.}
\vskip 0.1in
\resizebox{\linewidth}{!}{%
\begin{tabular}{@{}lccc@{}}
\toprule
\multirow{2}{*}{Dimensions} & Scenario & Empathetic & Empathetic \\
 & Understanding & Planning & Actions \\
\midrule
Action/Dialogue Assoc. & \ding{51} & \ding{51} & \ding{51} \\
Coherence              & \ding{51} & \ding{51} & \ding{51} \\
Emotional Communication & \ding{51} & \ding{51} & \ding{51} \\
Individual Understanding & \ding{51} & \ding{51} & \ding{51} \\
Emotion Regulation      & \ding{55}     & \ding{51} & \ding{51} \\
Helpfulness             & \ding{55}     & \ding{51} & \ding{51} \\
Adaptability            & \ding{55}     & \ding{51} & \ding{51} \\
Legality                & \ding{55}     & \ding{55}     & \ding{51} \\
\bottomrule
\end{tabular}
}
\label{tab:metric_usage}
\end{table}

\begin{table*}[ht]
\small
    \centering
    % \vskip 0.1in
    \caption{\textbf{Reference-based experiment results on scenario understanding and empathetic planning.} In scenario understanding, the model outputs a scenario description. In empathetic planning, the model outputs high-level planning. The input is the character's background, dialogue, and the video.
    We use the standard NLG metrics and compare the model's output with the ground truth. We find that GPT-4o performs the best on both scenario understanding and empathetic planning, suggesting the strongest ability to comprehend the empathetic need in scenarios and then plan responsively.}
    \vskip 0.1in
    \resizebox{\linewidth}{!}{%
    \begin{tabular}{@{}lcccccc|cccccc@{}}
    \toprule
    Task & \multicolumn{6}{c}{Scenario Understanding} & \multicolumn{6}{c}{Empathetic Planning} \\
    \cmidrule(lr){2-7} \cmidrule(lr){8-13}
    Metric & Bleu-1 & Bleu-4 & ROUGE-L & CIDEr & SPICE & BERTScore & Bleu-1 & Bleu-4 & ROUGE-L & CIDEr & SPICE & BERTScore \\
    \midrule
    LLaVA & 13.7 & 2.7 & 15.6 & 7.2 & 8.9 & 0.576 & 13.1 & 2.6 & 17.3 & 3.7 & 8.4 & 0.568 \\
    GPT-4-turbo & 14.1 & 3.1 & 20.4 & 1.6 & 10.1 & 0.612 & 25.7 & 6.9 & 23.5 & 14.9 & 14.5 & 0.621 \\
    GPT-4-vision & 15.2 & 3.3 & 21.4 & 3.1 & 12.1 & 0.615 & 25.9 & 6.4 & 23.4 & 15.5 & 11.8 & 0.625 \\
    GPT-4o & \textbf{19.1} & \textbf{5.3} & \textbf{23.7} & \textbf{8.8} & \textbf{14.8} & \textbf{0.622} & \textbf{30.8} & \textbf{12.0} & \textbf{26.1} & \textbf{25.9} & \textbf{16.7} & \textbf{0.641} \\
    \bottomrule
    \end{tabular}
    }
    \label{tab:L1_and_L2_results}
\end{table*}

\begin{table}[ht]
\small
    \centering
    % \vskip 0.1in
    \caption{\textbf{Reference-based experiment results on empathetic actions with multi-modal input.} The model outputs the grounded actions given the video, the character's information, and the dialogue. We use the Action Overlapping rate, TF-IDF, and LCS between the model's output and the ground truth. We find that GPT-4-vision performs the best on outputting the grounded actions, suggesting that these models are better at grounding to the simulated environment.}
    \vskip 0.1in
    \resizebox{\linewidth}{!}{%
    \begin{tabular}{@{}lccccc@{}}
    \toprule
    Metric & GPT-4o & GPT-4-turbo & GPT-4-vision & LLaVA & Qwen\\
    \midrule
    Overlap & 27.60 & 32.14 & \textbf{35.20} &  17.19 & 3.33\\
    TF-IDF & 21.03 & 24.76 & \textbf{27.69} &  12.09 & 1.85\\
    LCS & 25.17 & 28.92 & \textbf{29.58} & 15.21 &2.00 \\
    \bottomrule
    \end{tabular}
    }
    \label{tab:L3_results}
\end{table}

\section{Experiments}
In this section, we benchmark existing large language models (\ac{llm}s) and multimodal large language models (\ac{mllm}s) on EmpathyAgent. 
We also finetune Llama3-8B on EmpathyAgent, demonstrating its effectiveness in enhancing the empathetic behavior of embodied agents.

% \subsection{Benchmarking Results}
\subsection{Benchmarking Results on Reference-Based Metrics}
%bench mark models: GPT4o gpt4v gpt4turbo Llama 3 gpt3.5 qwen llava
%input level: video, action ground-truth
%results
% \subsubsection{Reference-Based Metrics}
We first provide benchmarking results on the reference-based metrics. To evaluate EmpathyAgent on different existing models, we use the most capable models publicly available: GPT-4-turbo, GPT-4-vision-preview, GPT-4o \cite{openai2023gpt4}, GPT-3.5 \cite{ouyang2022training}, Qwen (\texttt{qwen-vl-plus}) \cite{bai2023qwen} and LLaVA (\texttt{llava-13b}) \cite{liu2023visual}. With the temperature set to zero, we employ a consistent input prompt across all models. To facilitate efficient evaluation, we randomly select 100 samples from the test set, forming a testmini subset. Using our evaluation framework, we conduct experiments on this testmini subset to assess model performance. We provide additional quantitative results for other baselines in \Cref{sec:Additional_Baseline_Model}.

% \begin{table}[h]
% \small
%     \centering
%     \begin{tabular}{@{}llccccc@{}}
%     \toprule
%     Task & Metric & GPT-4o & GPT-4-turbo & GPT-4-vision & LLaVA \\
%     \midrule
%     \multirow{6}{*}{Scenario Understanding} 
%      & Bleu-1 & \textbf{19.1} & 14.1 & 15.2 & 13.7 \\
%      & Bleu-4 & \textbf{5.3} & 3.1 & 3.3 & 2.7 \\
%      & ROUGE-L & \textbf{23.7} & 20.4 & 21.4 & 15.6 \\
%      & CIDEr & \textbf{8.8} & 1.6 & 3.1 & 7.2 \\
%      & SPICE & \textbf{14.8} & 10.1 & 12.1 & 8.9 \\
%      & BERTScore & \textbf{0.622} & 0.612 & 0.615 & 0.576 \\
%     \midrule
%     \multirow{6}{*}{Empathetic Planning} 
%      & Bleu-1 & \textbf{30.8} & 25.7 & 25.9 & 13.1 \\
%      & Bleu-4 & \textbf{12.0} & 6.9 & 6.4 & 2.6 \\
%      & ROUGE-L & \textbf{26.1} & 23.5 & 23.4 & 17.3 \\
%      & CIDEr & \textbf{25.9} & 14.9 & 15.5 & 3.7 \\
%      & SPICE & \textbf{16.7} & 14.5 & 11.8 & 8.4 \\
%      & BERTScore & \textbf{0.641} & 0.621 & 0.625 &  0.568 \\
%     \midrule
%     \multirow{3}{*}{Empathetic Actions} 
%      & Overlap & 27.60 & 32.14 & \textbf{35.20} &  17.19 \\
%      & TF-IDF & 21.03 & 24.76 & \textbf{27.69} &  12.09 \\
%      & LCS & 25.17 & 28.92 & \textbf{29.58} & 15.21 \\
%     \bottomrule
%     \end{tabular}
%     \label{tab:combined_results}
% \end{table}

\begin{table*}[ht]
    \centering
    \small
    % \vskip 0.1in
    \caption{\textbf{Reference-based experiment results on empathetic actions with text-only input.} In this experiment, instead of directly observing the video, the model outputs the grounded actions given the \textbf{text description} of the video, the character's information, and the dialogue. We find that the instruction-finetuned model (\ie, Llama 3 IFT) on our benchmark attains the best performance on these metrics, suggesting that our benchmark can be used to boost empathetic actions in agents.}
    \vskip 0.1in
    % \resizebox{\linewidth}{!}{%
    \begin{tabular}{@{}lccccccc@{}}
    \toprule
    % Metric & GPT-4o & \makecell{GPT-4-\\turbo} & \makecell{GPT-4-\\vision} & \makecell{GPT-3.5-\\turbo} & Llama 3 & \makecell{Finetuned\\Llama 3}&\makecell{RLHF Finetuned\\Llama 3}\\
    Metric & GPT-4o & GPT-4-turbo & GPT-4-vision & GPT-3.5-turbo & Llama 3 & \makecell{Llama 3 IFT}&\makecell{Llama 3 RLHF}\\
    \midrule
      Overlap & 24.39 & 40.00 & 34.93 & 10.00 & 0.73 &\textbf{55.87} & 23.75\\
      TF-IDF & 18.32 & 31.56 & 28.29 & 9.33 & 0.41 &\textbf{47.34} & 18.41\\
      LCS &20.95 & 34.75& 30.67 & 9.67& 0.67&\textbf{49.83} & 20.35\\
    \bottomrule
    \end{tabular}
    % }
    \label{tab:L3_results_gt}
\end{table*}

\begin{table}[ht]
\small
\centering
% \vskip 0.1in
\caption{\textbf{Benchmarking results on the empathy evaluation framework.} GPT-4o outperforms LLaVA across all dimensions and evaluation steps, demonstrating superior performance in empathy-based metrics.}
\vskip 0.1in
\setlength{\tabcolsep}{2pt} 
\resizebox{\linewidth}{!}{%
\begin{tabular}{@{}lcccccc@{}}
\toprule
\multirow{3}{*}{Dimensions} & \multicolumn{2}{c}{Scenario} & \multicolumn{2}{c}{Empathetic} & \multicolumn{2}{c}{Empathetic}\\
 & \multicolumn{2}{c}{Understanding} & \multicolumn{2}{c}{Planning} & \multicolumn{2}{c}{Actions} \\
 & GPT-4o & LLaVA & GPT-4o & LLaVA & GPT-4o & LLaVA \\
\midrule
Action/Dialogue Assoc.      & 8.21 & 7.25 & 4.77 & 4.10 & 7.00 & 6.10 \\
Coherence      & 8.57 & 7.96 & 5.51 & 4.58 & 7.41 & 7.09 \\
Emotional Comm.   & 7.46 & 6.56 & 5.16 & 4.04 & 6.69 & 6.36 \\
Indiv. Understanding  & 6.91 & 6.64 & 4.63 & 3.92 & 5.69 & 5.39 \\
Emotion Regulation        & -    & -    & 7.09 & 4.96 & 8.43 & 7.91 \\
Helpfulness              & -    & -    & 5.76 & 4.95 & 8.08 & 7.35 \\
Adaptability             & -    & -    & 4.50 & 3.49 & 6.19 & 5.31 \\
Legality                 & -    & -    & -    & -    & 9.97 & 9.46 \\
\midrule
Overall Average         & \textbf{7.79} & 7.10 & \textbf{5.35} & 4.29 & \textbf{7.43} & 6.87 \\
\bottomrule
\end{tabular}
}
\label{tab:eight_dims}
\end{table}

% \begin{table*}[h]
% \small
%     \centering
%     \caption{\textbf{Combination results of experiments benchmarking models on reference-free metrics.} GPT-4o outperforms LLaVA across all dimensions and challenges, suggesting that GPT-4o consistently exhibits superior capabilities in empathy-based metrics.}
%     \vspace{0.2cm}
%     % \resizebox{\linewidth}{!}{%
%     \begin{tabular}{@{}lcccccc@{}}
%     \toprule
%     \multirow{2}{*}{Dimensions} & \multicolumn{2}{c}{Scenario Understanding} & \multicolumn{2}{c}{Empathetic Planning} & \multicolumn{2}{c}{Empathetic Actions} \\
%      & GPT-4o & LLaVA & GPT-4o & LLaVA & GPT-4o & LLaVA \\
%     \midrule
%     Action and Dialogue Association & 8.21 & 7.25 & 4.77 & 4.10 & 7.00 & 6.10 \\
%     Coherence & 8.57 & 7.96 & 5.51 & 4.58 & 7.41 & 7.09 \\
%     Emotional Communication & 7.46 & 6.56 & 5.16 & 4.04 & 6.69 & 6.36 \\
%     Individual Understanding & 6.91 & 6.64 & 4.63 & 3.92 & 5.69 & 5.39 \\
%     Emotion Regulation & - & - & 7.09 & 4.96 & 8.43 & 7.91 \\
%     Helpfulness & - & - & 5.76 & 4.95 & 8.08 & 7.35 \\
%     Adaptability & - & - & 4.50 & 3.49 & 6.19 & 5.31 \\
%     Legality & - & - & - & - & 9.97 & 9.46 \\
%     Overall Average & \textbf{7.79} & 7.10 & \textbf{5.35} & 4.29 & \textbf{7.43} & 6.87 \\
%     \bottomrule
%     \end{tabular}
%     % }
%     \label{tab:eight_dims}
% \end{table*}

\begin{figure*}[t]
% \vspace{-0.2cm}
    \centering
    \includegraphics[width=1\linewidth]{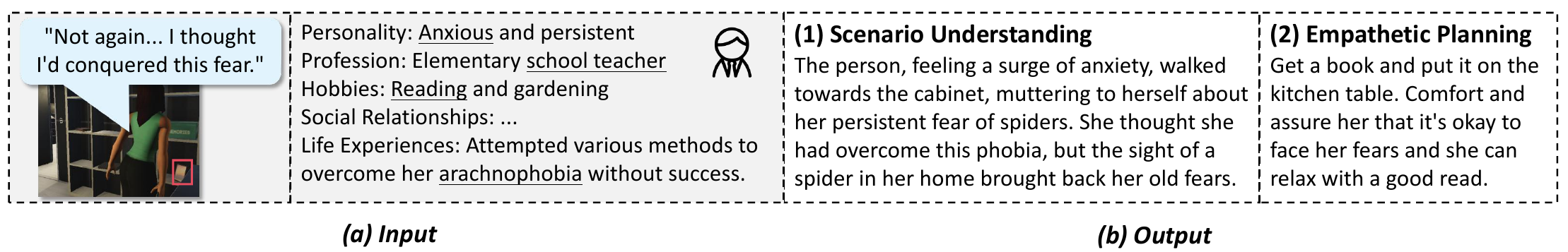}
    % \vskip 0.1in
    \caption{\textbf{Qualitative results of GPT-4o.} We test the scenario understanding and empathetic planning capability of GPT-4o. We find that GPT-4o has strong capabilities in empathetic scenario understanding and high-level empathetic planning.}
    % \vskip 0.1in
    \label{fig:gpt4o_qualitative}
\vspace{-0.4cm}
\end{figure*}
\textbf{Scenario Understanding} In this experiment, we input videos, character information, and dialogue. The model outputs the scenario description based on its understanding.
For the GPT models, we input one frame for every sequential five frames. For LLaVA, we input the middle frame only as LLaVA doesn't support multi-image input. 
We use a human-annotated example to prompt the model to generate the scenario description. We then compare it with the ground truth and report the NLG metrics. The results are presented in \Cref{tab:L1_and_L2_results}. We find that GPT-4o performs the best, indicating its potential to understand the causes and underlying emotions of a scenario.\\
\textbf{Empathetic Planning}
In goal inference experiments, we input videos, character information, and dialogue of the character to the models like scenario understanding. The results are presented in \Cref{tab:L1_and_L2_results}, where we find that GPT-4o performs the best on high-level empathetic planning.
Then we present a qualitative result of scenario understanding and empathetic planning of GPT-4o in \Cref{fig:gpt4o_qualitative}. The model demonstrates good scene understanding and planning abilities. \\
\textbf{Empathetic Action}
In empathetic action experiments, we experiment in two settings: 
\textbf{(i) Video Scenario Input} 
We input the scenario video into the model and evaluate the output actions generated by GPT models, Llava, and Qwen.
The results are shown in \Cref{tab:L3_results}, where GPT-4-vision-preview performs the best at outputting grounded actions. 
Although GPT-4o performs well on scene understanding and high-level planning, it still needs improvement in outputting grounded action sequences. 
\textbf{(ii) Text Scenario Input} 
We use the text-formed description of the video and test it on both multi-modal models and LLMs. We present the results in \Cref{tab:L3_results_gt}. Among the pre-trained models, GPT-4-turbo performs the best. 
% More qualitative results are provided in Appendix.D.
\subsection{Benchmarking Results on Reference-Free Metrics}
% \subsubsection{Reference-Free Metrics}
We also benchmark GPT-4o and LLaVA (llava-13b) using our reference-free metrics. We assess them across eight dimensions that span the three challenges of empathy response. The results are shown in \Cref{tab:eight_dims}. We find that GPT-4o consistently outperforms LLaVA across all dimensions and challenges, demonstrating its stronger capabilities in alignment with empathy-based principles. Notably, both models perform weakest in Individual Understanding and Adaptability, indicating that improvements in these aspects could advance future research aimed at enhancing empathetic abilities in models.

\subsection{Trained Empathetic Agent Results}
% \begin{figure*}[ht]
%     \centering
%     \includegraphics[width=0.8\linewidth]{images/llama3_qualitative.pdf}
%     \vskip 0.1in
%     \caption{\textbf{Qualatitive comparison between Llama-3-8B and Llama-3-8B instruction-finetuned on our benchmark.} The pretrained Llama8B often struggles to understand the actions and chooses not to take any actions in most cases. The dialogue is also simple and not empathetically responsive. After finetuning, Llama3-8B is able to conduct a series of empathetic actions and output a dialogue that is more empathetically responsive.}
%     \vskip 0.1in
%     \label{fig:qualitative}
% % \vspace{-0.2cm}
% \end{figure*}
\begin{figure}[ht]
\vspace{-0.2cm}
    \centering
\includegraphics[width=1\linewidth]{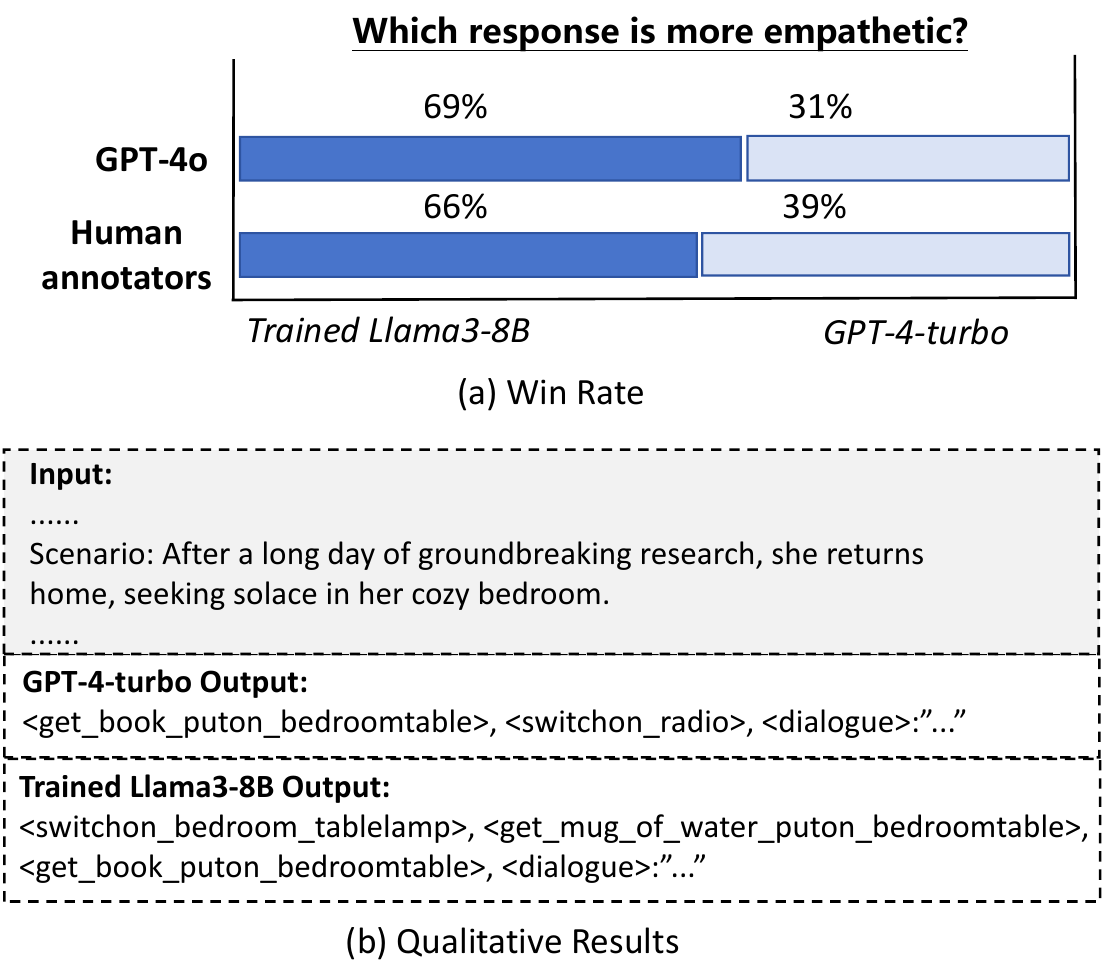}
    % \vskip 0.1in
    \vspace{-0.2cm}
    \caption{\textbf{Comparison between GPT-4 and instruction tuned Llama3-8B.} We sample 10 pairs of data and report the GPT win rate and human win rate. Specifically, we ask either GPT/human annotator to choose which response is more empathetic. We find that instruction-finetuned Llama3-8B outperforms GPT-4-turbo with significantly fewer parameters, suggesting that the benchmark can be potentially leveraged to build a powerful empathetic agent.}
    % \vskip 0.1in
    \label{fig:win_rate}
\vspace{-0.2cm}
\end{figure}
\paragraph{Instruction-Finetuned Empathetic Agent}
In this experiment, we train Llama3-8B on our training set using instruction finetuning. 
For the training part, we use the text-formed input actions, the dialogue, and the character's background as input and directly finetuned the action-level response. For the testing part, we let the model directly output the low-level actions and conduct action-level testing. 
The quantatitive results are shown in \Cref{tab:L3_results}, suggesting that such training boosts empathetic behavior. 
% As shown in \Cref{fig:qualitative}, 
Before training, the Llama-8b model almost cannot conduct any empathetic actions. In most cases, it chooses not to conduct any empathetic actions but only outputs a short dialogue. After training, the model is able to conduct a series of empathetic actions, and the output dialogue is also more empathetic. More qualitative results are shown in \Cref{sec:Additional_Qualitative_Results}.

We also compare the trained Llama3-8B with the GPT-4 model. We first evaluate with the reference-based metrics and show the results in \Cref{tab:L3_results}. By using EmpathyAgent for training, the instruct-finetuned Llama3 with only 8B parameters outperforms GPT-4 on these metrics.
We then conduct an evaluation of human preference and GPT preference and report the GPT-win rate and human-win rate. Specifically, for the GPT-win-rate, we provide character information, ground truth action list, scenario description, dialogue, and the two responses generated by GPT-4-turbo and the instruct-finetuned Llama3 and let GPT-4o choose the more empathetic response. For the human win rate, we randomly sample 10 pairs of data from our test set and ask 10 human annotators to choose the more empathetic one. 
The results are shown in \Cref{fig:win_rate}, instruction-finetuned Llama3-8B outperforms GPT-4-turbo on both GPT-4o and human preference, suggesting that EmpathyAgent can be used in future studies to effectively train empathetic agents. 
\vspace{-5pt}
\paragraph{RLHF Empathetic Agent}
Lastly, we use the paired data and RLHF technique to train Llama3-8B. We first use the paired preference data to train a reward model and then train the Llama3 model using the reward as feedback. The results are shown in \Cref{tab:L3_results_gt}
, which indicates that RLHF training is also capable of boosting empathetic performance, but is not as effective as instruction finetuning. We believe this could be due to insufficient training of the reward model. We will work on developing a more robust reward model to assign scores for empathetic responses.

% \section{Conclusion and Limitation}
\section{Conclusion}
In this work, we introduce EmpathyAgent, the first benchmark specifically designed for evaluating and benchmarking the empathetic actions of embodied agents. 
% Embodied agents are required to perform actions based on their understanding of both the visual scene and human emotions.
Our benchmark contains 10,000 samples, encompassing multimodal inputs and corresponding empathetic task planning sequences across diverse scenarios. The benchmark generation method mimics the human empathy process and can be scaled up automatically. 
Furthermore, we design three challenges for the empathy response and develop a systematic evaluation framework with reference-based and reference-free metrics, conducting comprehensive evaluations on the most capable models. Finally, we finetune Llama3-8B on our benchmark, demonstrating its effectiveness in enhancing the empathetic behavior of embodied agents. EmpathyAgent is the first to advance the study of building embodied agents that provide empathetic support to humans.
% Regarding limitations, we currently use a large-sized \ac{llm} to evaluate our EmpathyAgent benchmark, but the inference speed is relatively slow. To improve practicality, we plan to use smaller-sized \ac{llm}s or explore quantizing and compressing the model in the future. Meanwhile, we will add more human-labeled data to provide additional choices made by humans for empathetic responses.  

% arxiv
% \section*{Acknowledgements}
% This work was supported by the National Natural Science Foundation of China (62476011).

\section*{Impact Statement}
This paper presents work whose goal is to advance the field of Embodied Agents. There are many potential societal consequences of our work, none which we feel must be specifically highlighted here.

\section*{Acknowledgements}
This work was supported by the National Natural Science Foundation of China (62476011).

\bibliography{example_paper}
\bibliographystyle{icml2025}

%%%%%%%%%%%%%%%%%%%%%%%%%%%%%%%%%%%%%%%%%%%%%%%%%%%%%%%%%%%%%%%%%%%%%%%%%%%%%%%
%%%%%%%%%%%%%%%%%%%%%%%%%%%%%%%%%%%%%%%%%%%%%%%%%%%%%%%%%%%%%%%%%%%%%%%%%%%%%%%
% APPENDIX
%%%%%%%%%%%%%%%%%%%%%%%%%%%%%%%%%%%%%%%%%%%%%%%%%%%%%%%%%%%%%%%%%%%%%%%%%%%%%%%
%%%%%%%%%%%%%%%%%%%%%%%%%%%%%%%%%%%%%%%%%%%%%%%%%%%%%%%%%%%%%%%%%%%%%%%%%%%%%%%
\newpage
\appendix
\onecolumn
\section{Appendix}
\subsection{Overview}
% In this section, we provide an overview of our supplementary materials listed below. 
We organize our supplementary material as follows.
% We first introduce additional dataset details, which include additional examples, the character pool and the input actions pool in our dataset, and the prompts we used to create the dataset. Then, we introduce our implementation details, which include baseline details and training details. Finally, we provide some additional results, including quantitative and qualitative results.
\begin{itemize}
\item Additional Related Work
\item Benchmark Details
\begin{itemize}
    \item Data Statistics
    \item Additional Examples 
    \item Character Pool Details
    \item Input Actions Pool Details
    \item Labels of Empathetic Action Sequences
    \item Prompt Details
\end{itemize}
\item Metric Design Details
\item Additional Quantitative Results
\begin{itemize}
    \item Implementation Details
    \begin{itemize}
        \item Training Details
         \item Details of Metrics in the Empathetic Action Process
    \end{itemize}
    \item Additional Baseline Model
\end{itemize}
\item Additional Qualitative Results
\begin{itemize}
        \item Instruct Finetuned Empathetic Agent
         \item RLHF Empathetic Agent
\end{itemize}
\end{itemize}

\subsection{Additional Related Work}
\textbf{\ac{llm} as Social Agent} 
\acp{llm} have shown impressive out-of-box common sense reasoning abilities \cite{kim2023soda, west2021symbolic} and can even have personalities \cite{jiang2023evaluating}.
Recent works have used \acp{llm} as generative agents \cite{park2023generative} that can plan, reason, and interact in a simulated environment. 
Sotopia designs various characters and studies their social intelligence \cite{zhou2023sotopia}. \cite{liu2023training} studies training \acp{llm} to effectively learn from simulated social interactions.
CAMEL \cite{li2023camel} studies the collaborative problem-solving of \acp{llm}. 
However, these agents are not embodied. They are largely limited to dialogues and cannot be applied to a grounded environment to perform executable actions. We bridge this gap to let these agents engage in a simulated robotic navigation environment where these agents need to interact with objects and perform grounded actions.

\subsection{Benchmark Details}\label{sec:dataset_details}

\subsubsection{Data Statistics}
We provide the key statistics of EmpathyAgent in \Cref{tab:dataset_statistics}. Our benchmark contains 10k samples, including 100 different characters and 20 different input action videos.
\begin{table}[ht]
\small
    \centering
    \vskip 0.1in
    \caption{\textbf{Key statistics in EmpathyAgent.} Our benchmark contains 10k samples, including 100 different characters and 20 different input action videos.}
    \vskip 0.1in
    \begin{tabular}{@{}lc@{}}
    \toprule
    Statistic & Number \\
    \midrule
    Total Data Points & 10k \\
    Characters & 100 \\
    Input Action-Video & 20 \\
    Scenarios and Dialogues per Character-Action pair & 5 \\
    Empathy Response per Data Point & 2 \\
    Optional Action Space for Output & 50 \\
    Average Length of Action-Video & 16.28s \\
    Max Length of Action-Video & 24.60s \\
    Min Length of Action-Video & 9.40s \\
    \bottomrule
    \end{tabular}
    \label{tab:dataset_statistics}
\end{table}

\subsubsection{Additional Examples}\label{sec:additional_examples}
\begin{figure}
    \centering
\includegraphics[width=0.8\linewidth]{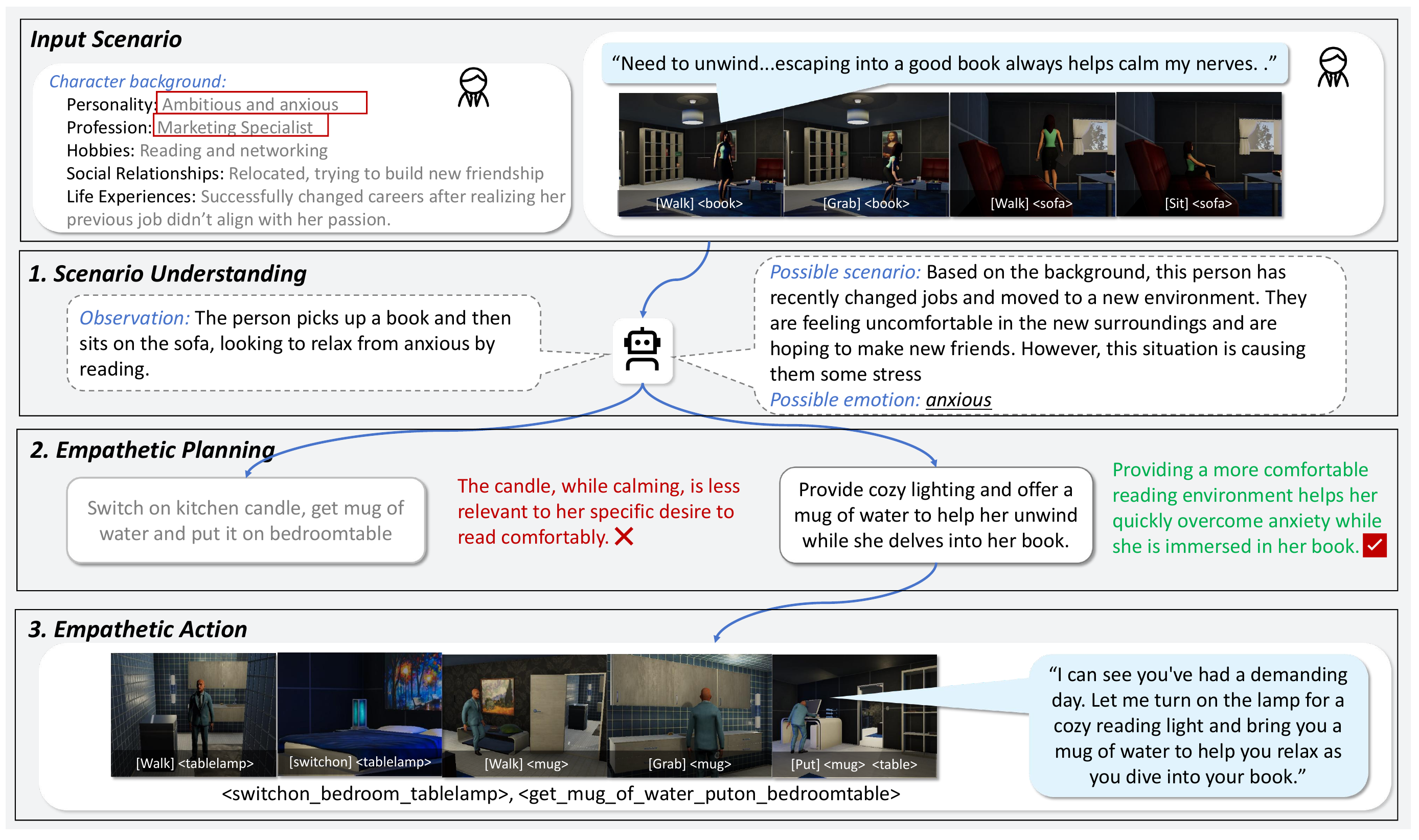}
\vskip 0.1in
    \caption{\textbf{Example of the EmpathyAgent.} In this example, the scenario contains an ambitious and anxious person who is looking for a book. The embodied agent first perceives the scenario and understands that the person has just moved to a new environment and is likely anxious at this point. Based on this understanding, the embodied agent comes up with a plan to provide a comfortable environment for this person. So the embodied agent takes the action to switch on the bedroom table lamp and get a mug of water to put on the bedroom table.}
    \vskip 0.1in
    \label{fig:dataset1}
\end{figure}

\begin{figure}
    \centering
    \vskip 0.1in
\includegraphics[width=0.8\linewidth]{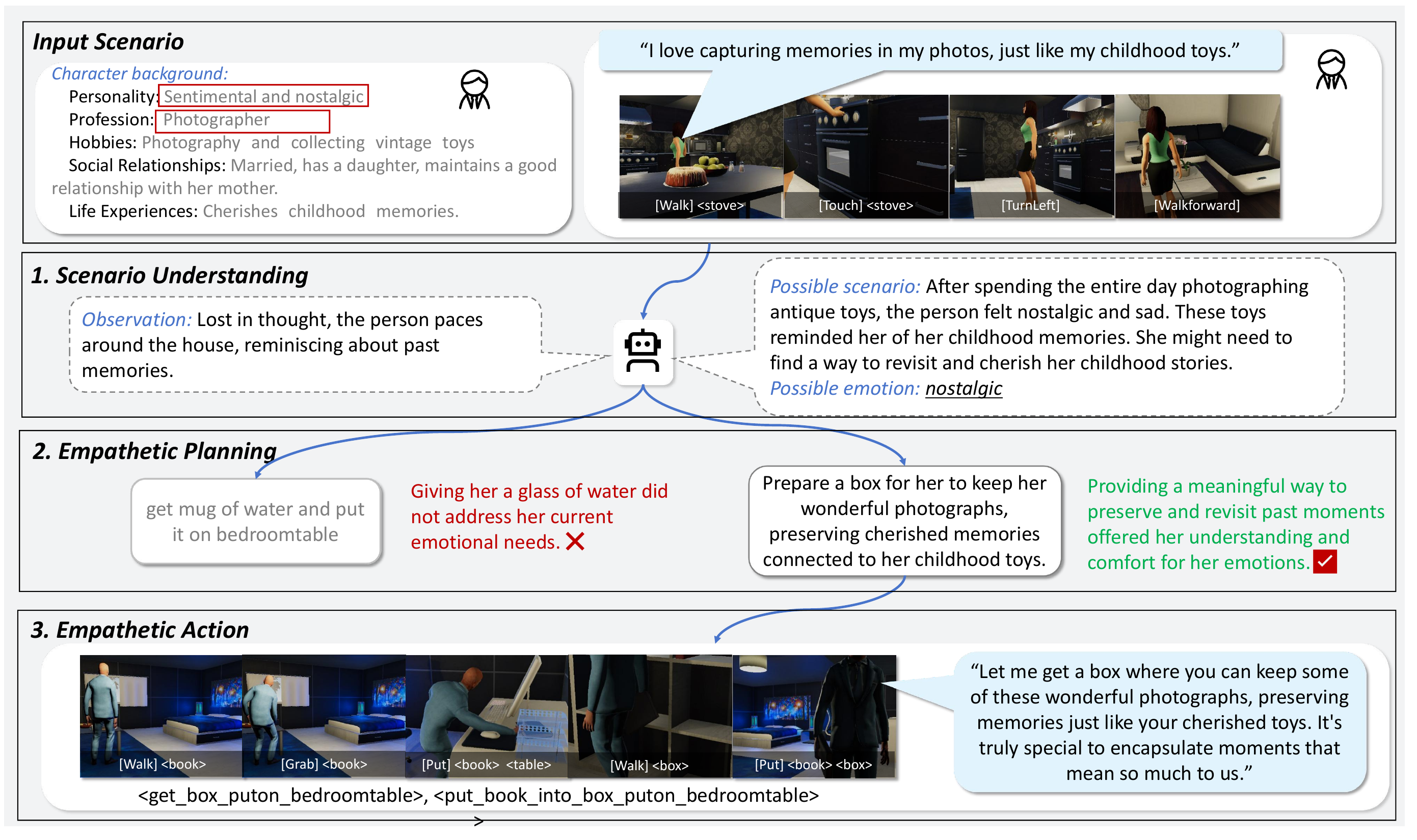}
\vskip 0.1in
     \caption{\textbf{Example of our Benchmark.} In this example, the scenario contains a sentimental and nostalgic person who is looking through past photos. The embodied agent first perceives the scenario and understands that the person has recalled her good old memories and is likely nostalgic at this point. Based on this understanding, the embodied agent comes up with a plan to help the person preserve her memories and comfort her. So the embodied agent gets her a box and comforts her.}
    \label{fig:dataset2}
\end{figure}

We provide additional examples of our benchmark as shown in \Cref{fig:dataset1,fig:dataset2}. 
% two more examples of our dataset in \Cref{dataset1} and \Cref{dataset2}
\subsubsection{Character Pool Details}
\begin{figure}
    \centering
\includegraphics[width=0.8\linewidth]{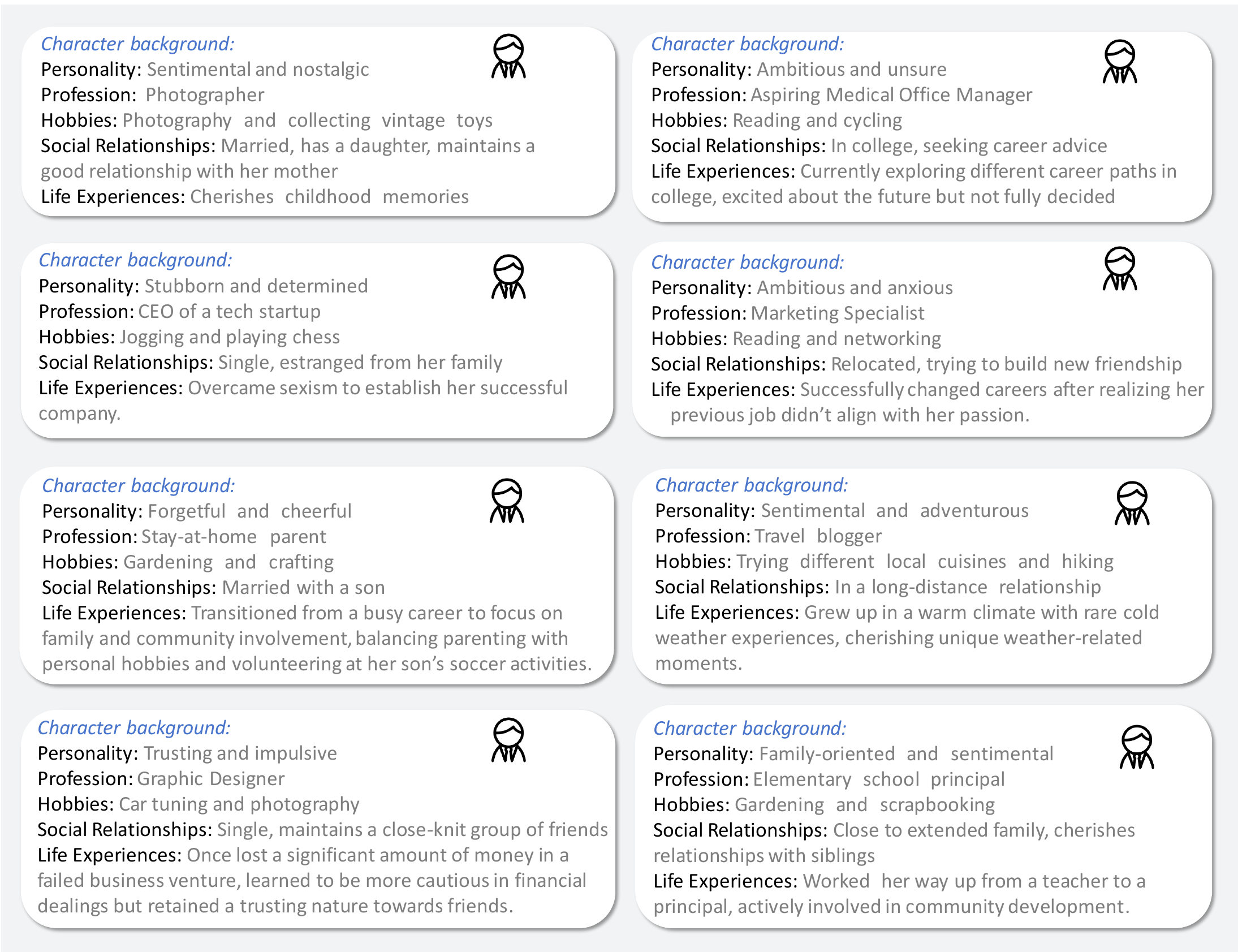}
    \vskip 0.1in
    \caption{\textbf{Examples of our Character Pool.} Each character contains a unique personal file, including personality, profession, hobbies, social relationships, and life experiences.}
    \vskip 0.1in
    \label{fig:character_pool}
\end{figure}

% total: 100 different characters
% in \Cref{character_pool}: 8 examples
We provide details of our character pool as shown in \Cref{fig:character_pool}. Each character contains a unique personal file, including personality, profession, hobbies, social relationships, and life experiences.

\subsubsection{Input Actions Pool Details}
\begin{figure}
    \centering
\includegraphics[width=0.8\linewidth]{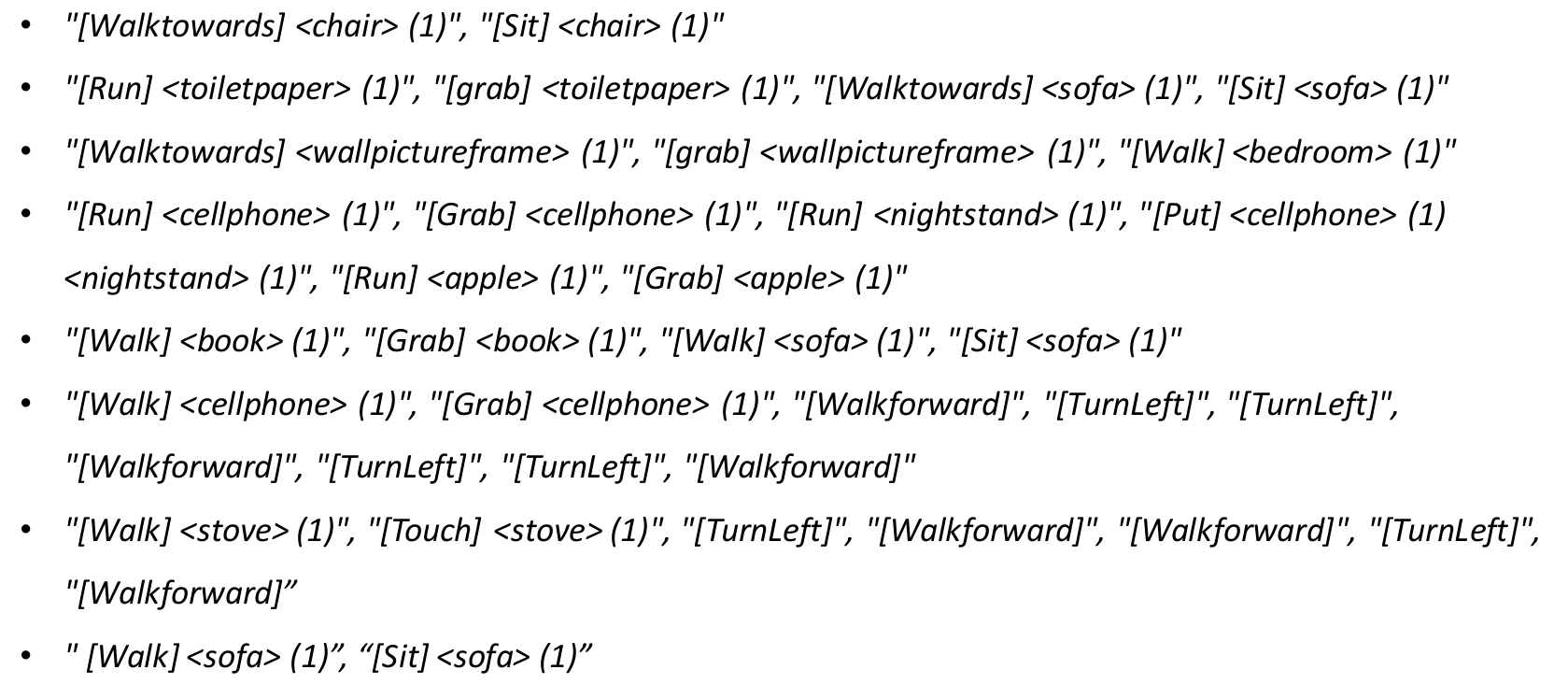}
\vskip 0.1in
    \caption{\textbf{Examples of the input action Pool.} We present examples of the input actions in VirtualHome, this can be further rendered into a video of a person performing these actions sequentially.}
    \vskip 0.1in
    \label{fig:action_list}
\end{figure}

We provide examples of our input actions pool as shown in \Cref{fig:action_list}. This contains a sequence of legal actions and can be further rendered into a video.
% Total: 20 action lists that can be rendered in virtualhome
% in \Cref{action_list}: 8 examples

\subsubsection{Labels of Empathetic Action Sequences}
In the Empathy Response Generation process, we generate empathetic action sequences and create labels for each of them. In \Cref{fig:labels}, we present several examples of the labels. 

\subsubsection{Prompt Details} 
\begin{figure}
    \centering
    \includegraphics[width=1\linewidth]{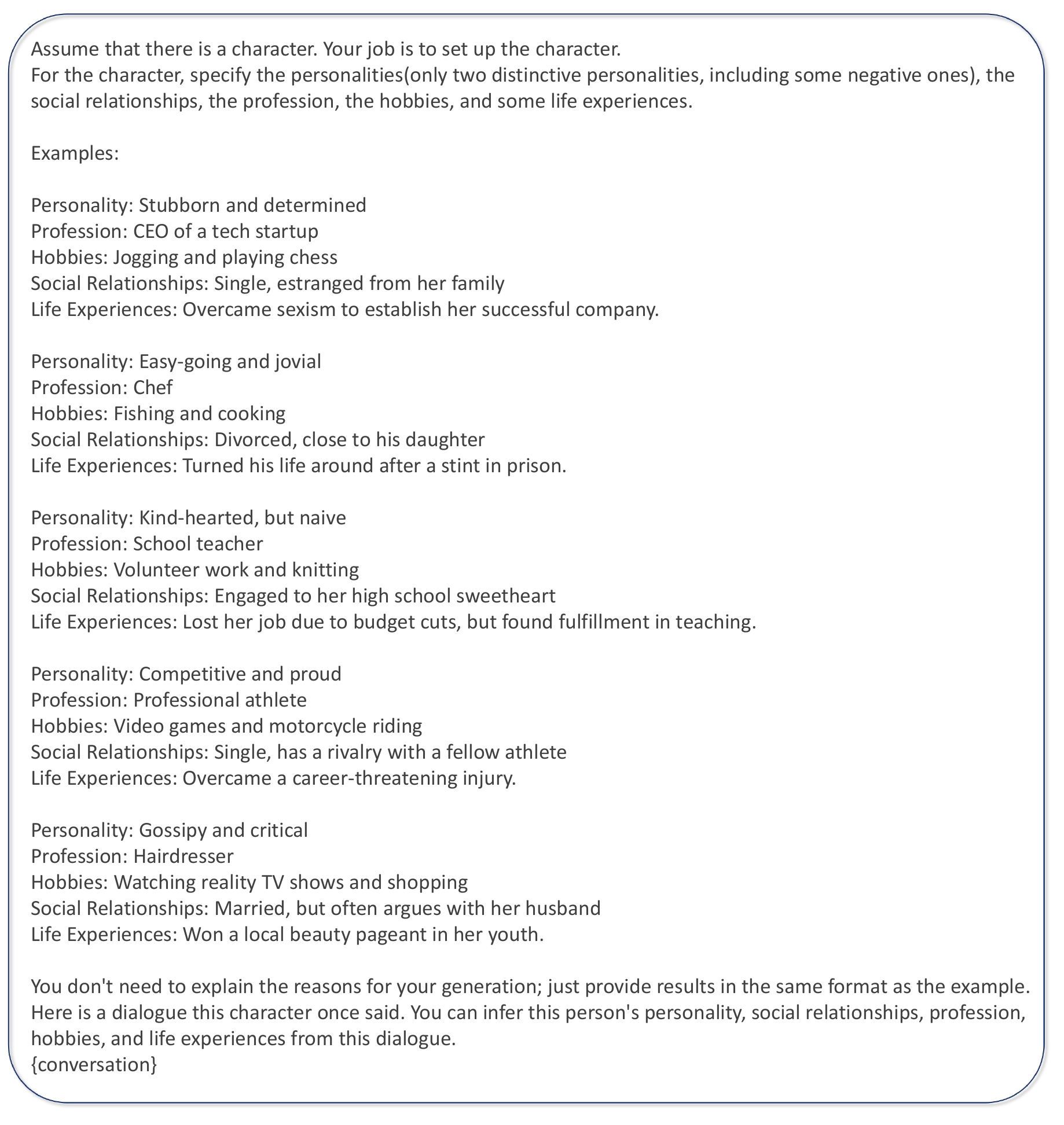}
    \vskip 0.1in
    \caption{\textbf{Prompt for generating the character pool}. The model is given 5 in-context examples and a data sample from EmpathyDialogue, then creates a new character profile.}
    \vskip 0.1in
    \label{fig:prompt_character}
\end{figure}

We provide the prompt we used to generate the benchmark.
\paragraph{Character Pool Generation}
The prompt we used to generate the character profiles is shown in \Cref{fig:prompt_character}. For each API call, we randomly sample a data point from the EmpatheticDialogues ~\cite{rashkin-etal-2019-towards} to fill the {conversation} field. This enhances the diversity and encourages the model to draw inspiration from dialogues that contain empathetic cues as shown in previous works~\cite{zhou2023sotopia}.
We use 5 in-context examples in this prompt.

\begin{figure}
    \centering
    \includegraphics[width=1\linewidth]{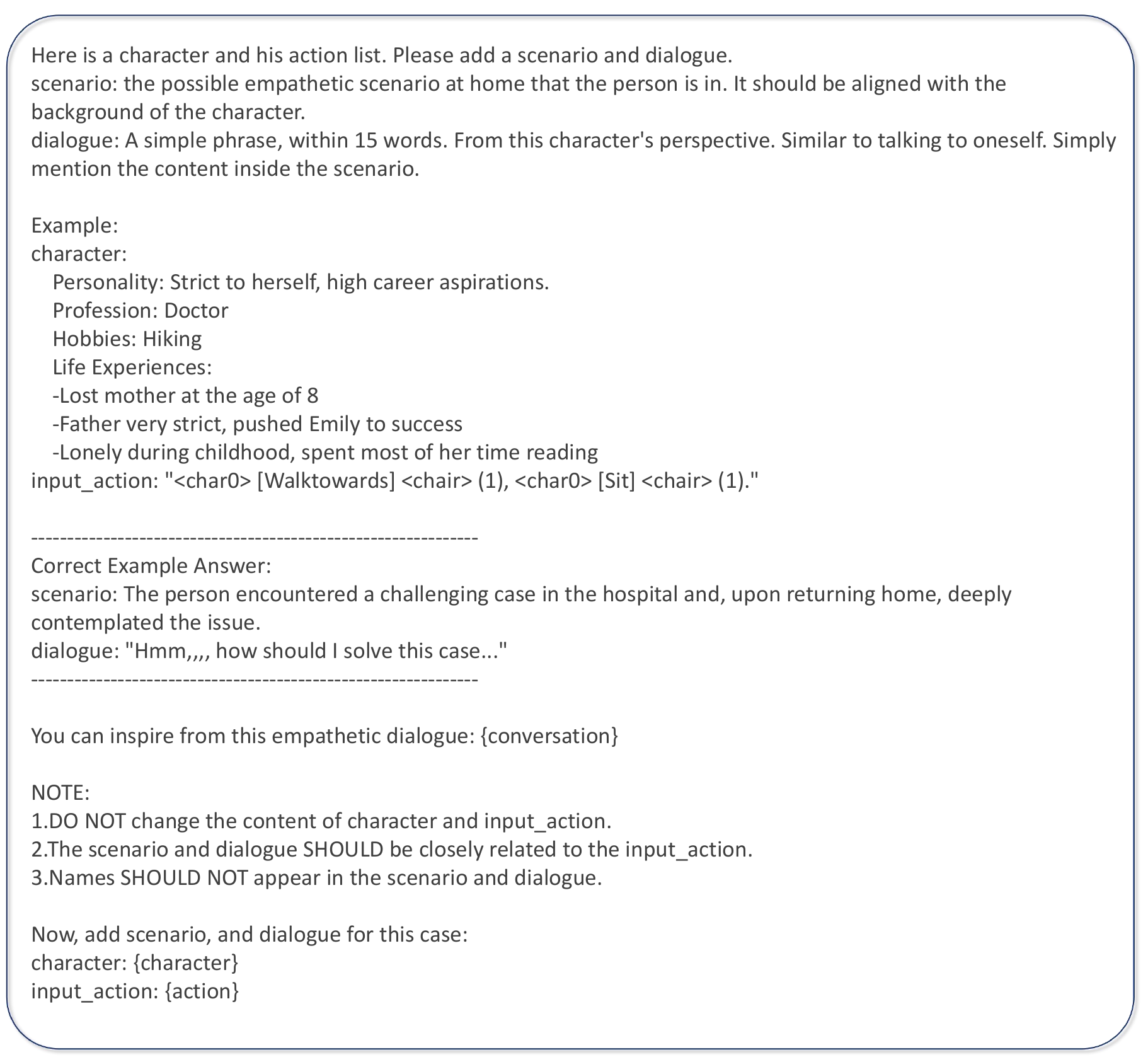}
    \vskip 0.1in
    \caption{\textbf{Prompt for creating the scenario.} Given the character's profile and input actions, the model outputs a scenario description and also the character's dialogue under this scenario.}
    \vskip 0.1in
    \label{fig:prompt_scenario}
\end{figure}

\begin{figure}
    \centering
\includegraphics[width=0.75\linewidth]{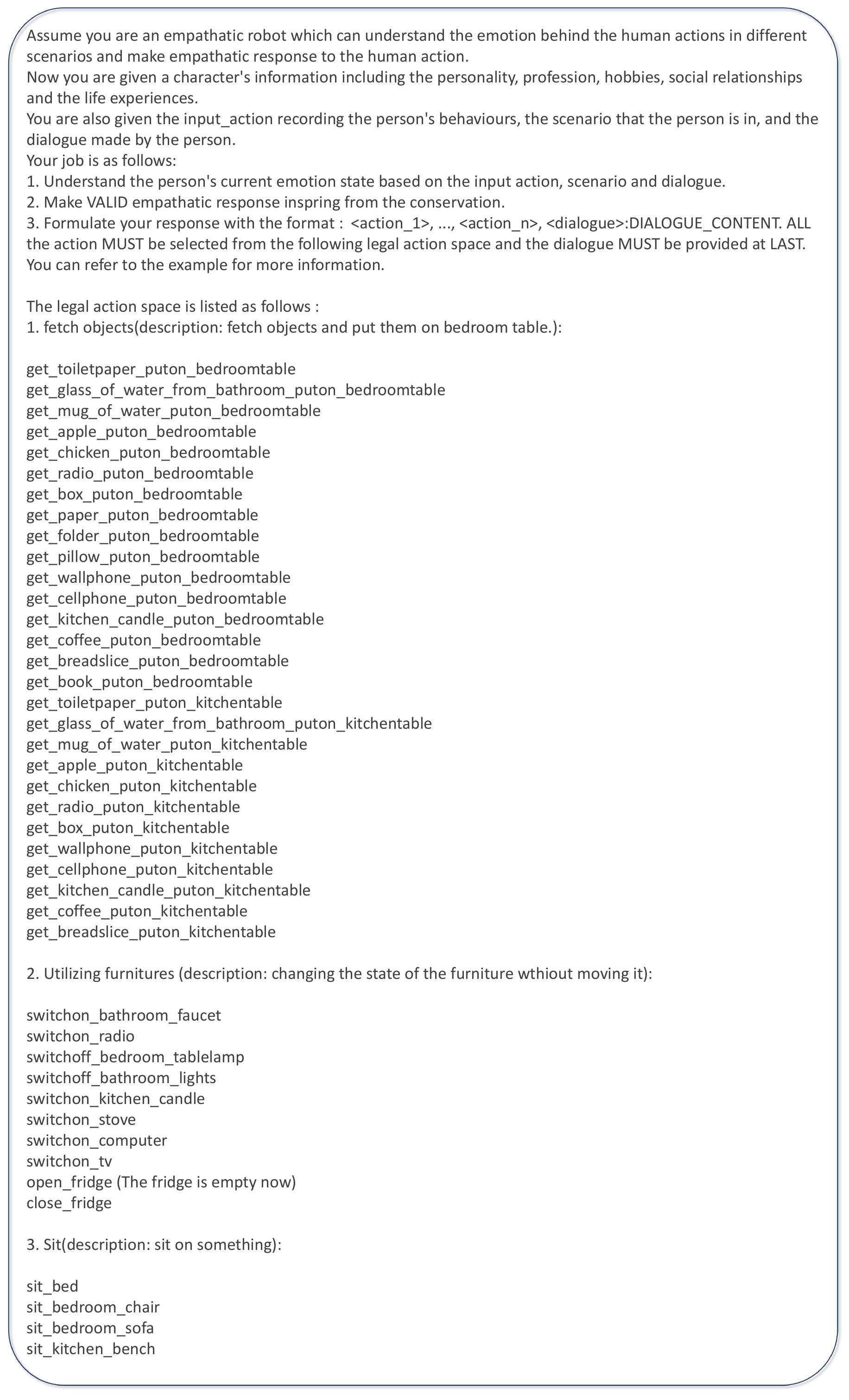}
    \vskip 0.1in
    \caption{\textbf{Prompt for generating the empathetic actions.}}
    \vskip 0.1in
    \label{fig:prompt_goal1}
\end{figure}

\begin{figure}
    \centering
    \includegraphics[width=0.8\linewidth]{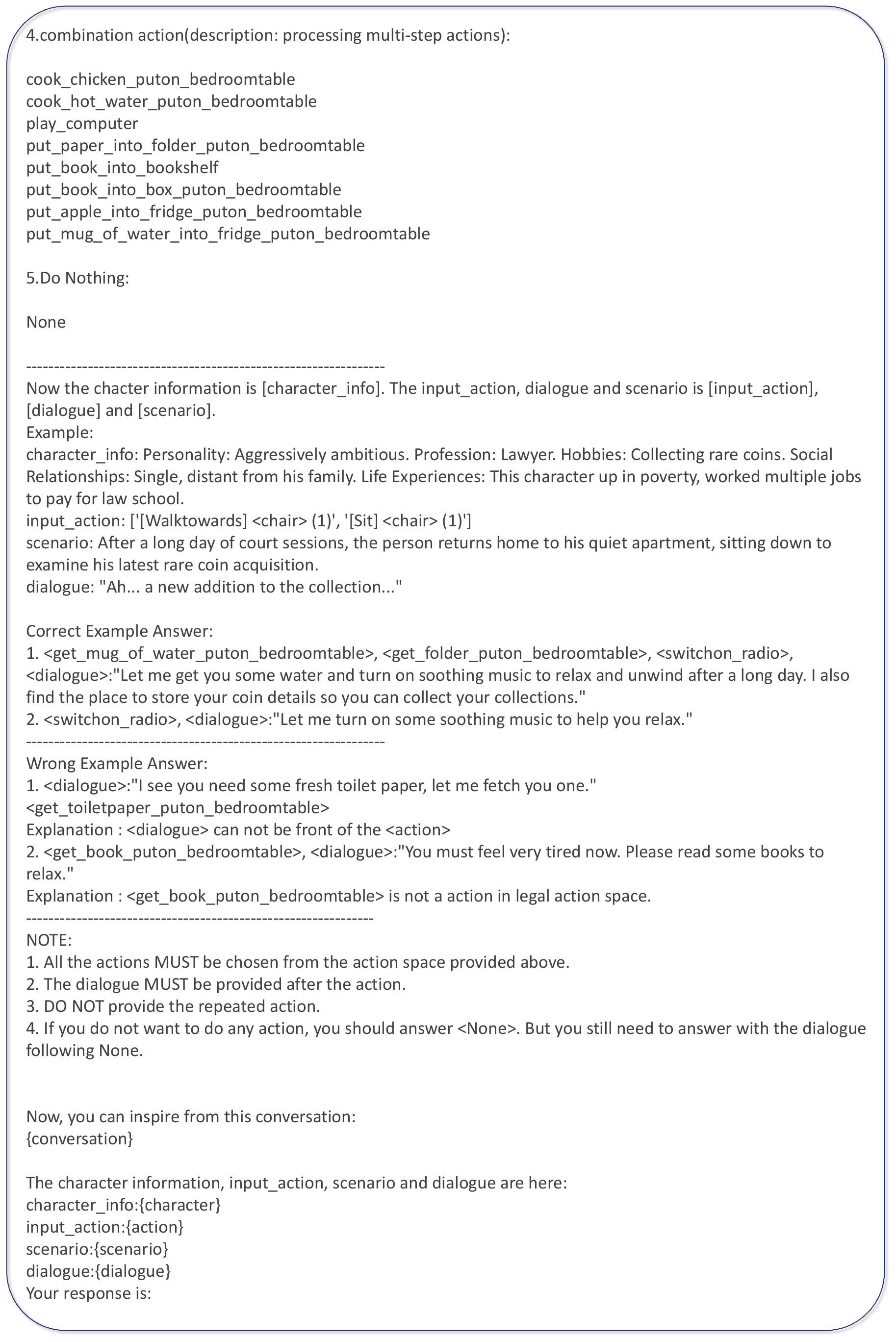}
    \vskip 0.1in
    \caption{\textbf{Prompt for generating the empathetic actions.}}
    \vskip 0.1in
    \label{fig:prompt_goal2}
\end{figure}

\begin{figure}
    \centering
    \includegraphics[width=0.65\linewidth]{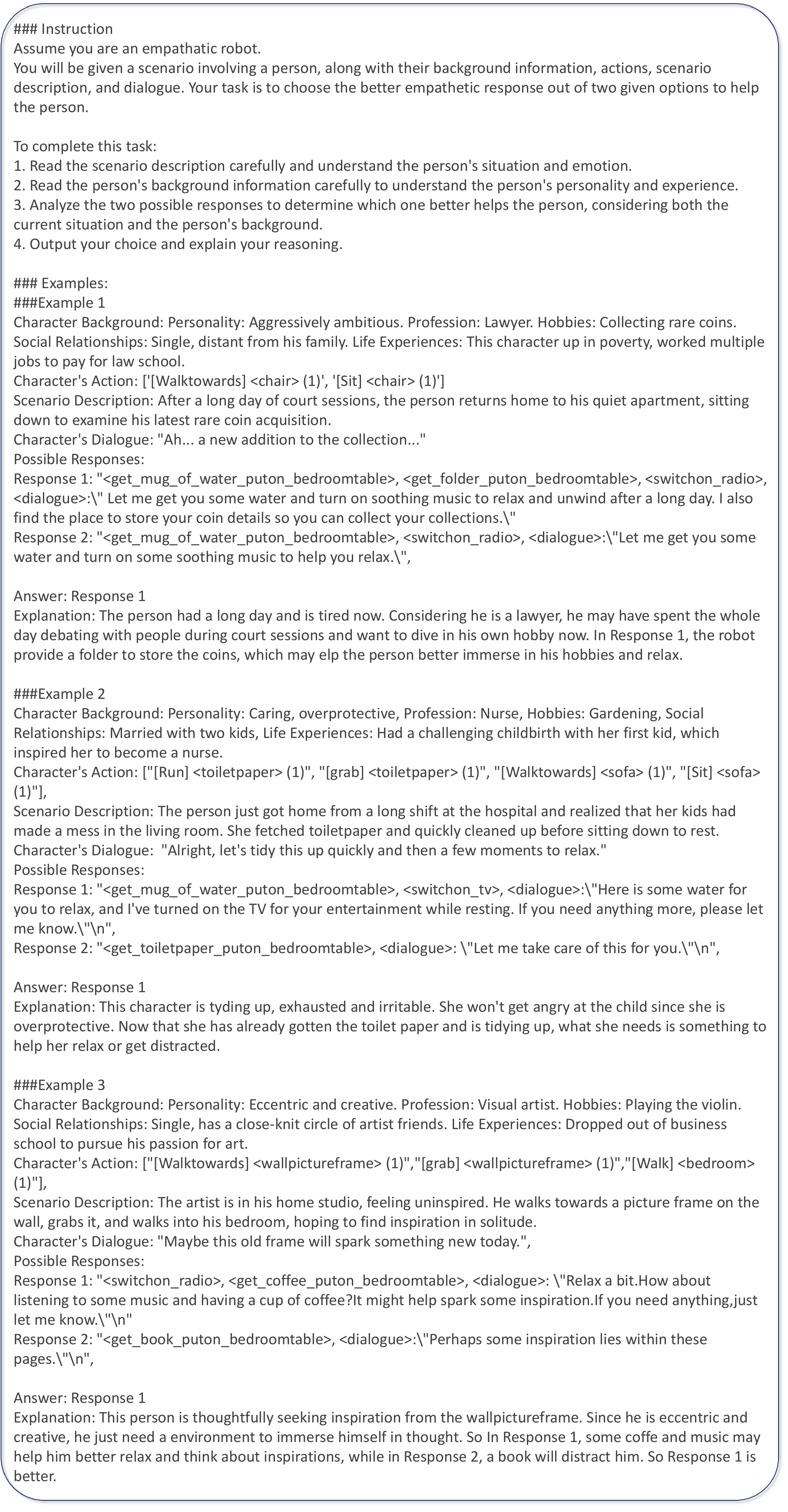}
    \vskip 0.1in
    \caption{\textbf{Prompt for selecting the more empathetic response and providing an explanation.}}
    \vskip 0.1in
    \label{fig:prompt_rank1}
\end{figure}

\begin{figure}
    \centering
    \includegraphics[width=0.65\linewidth]{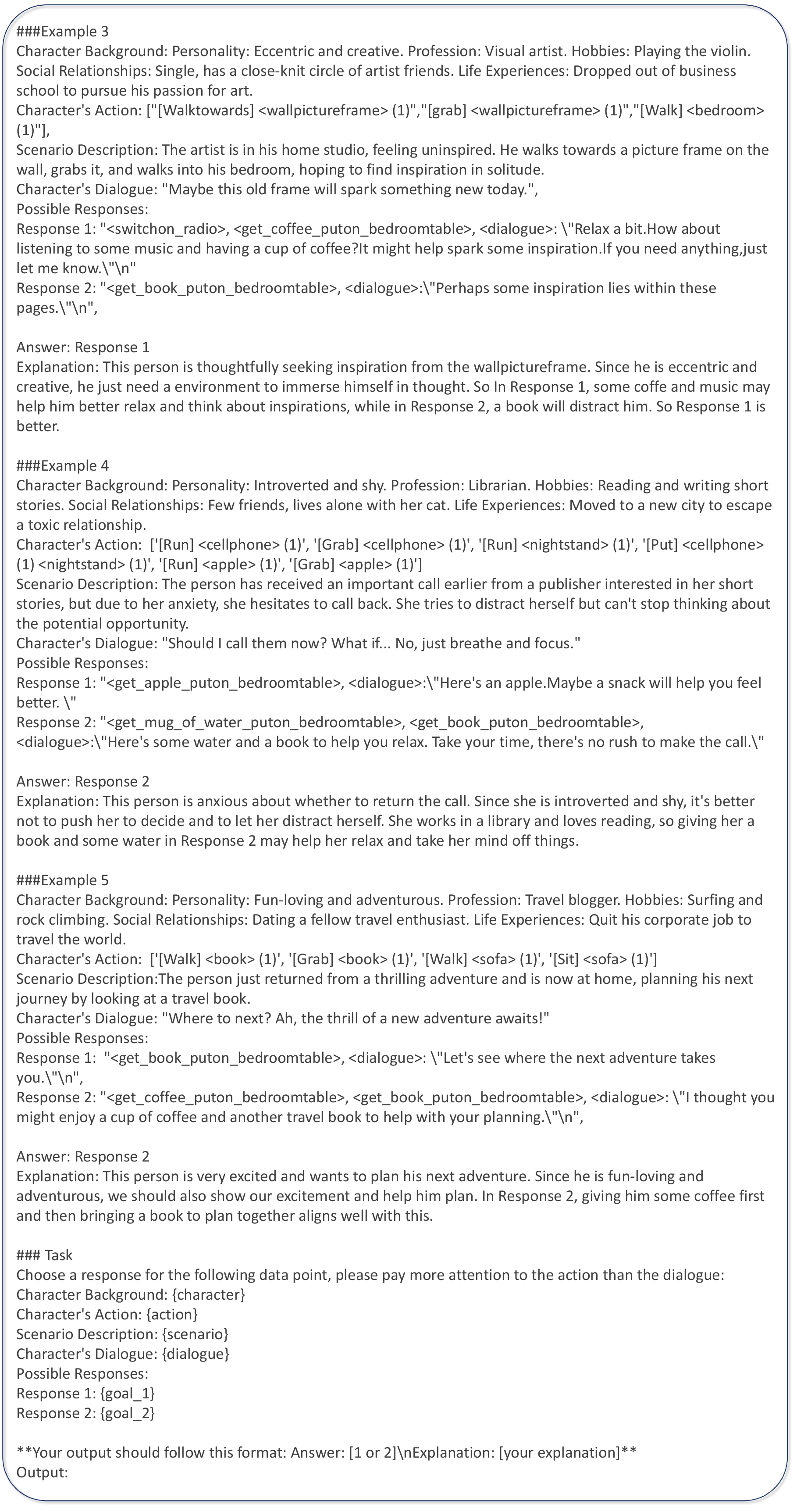}
    \vskip 0.1in
    \caption{\textbf{Prompt for selecting the more empathetic response and providing an explanation.}}
    \vskip 0.1in
    \label{fig:prompt_rank2}
\end{figure}

\begin{figure}
    \centering
    \includegraphics[width=0.85\linewidth]{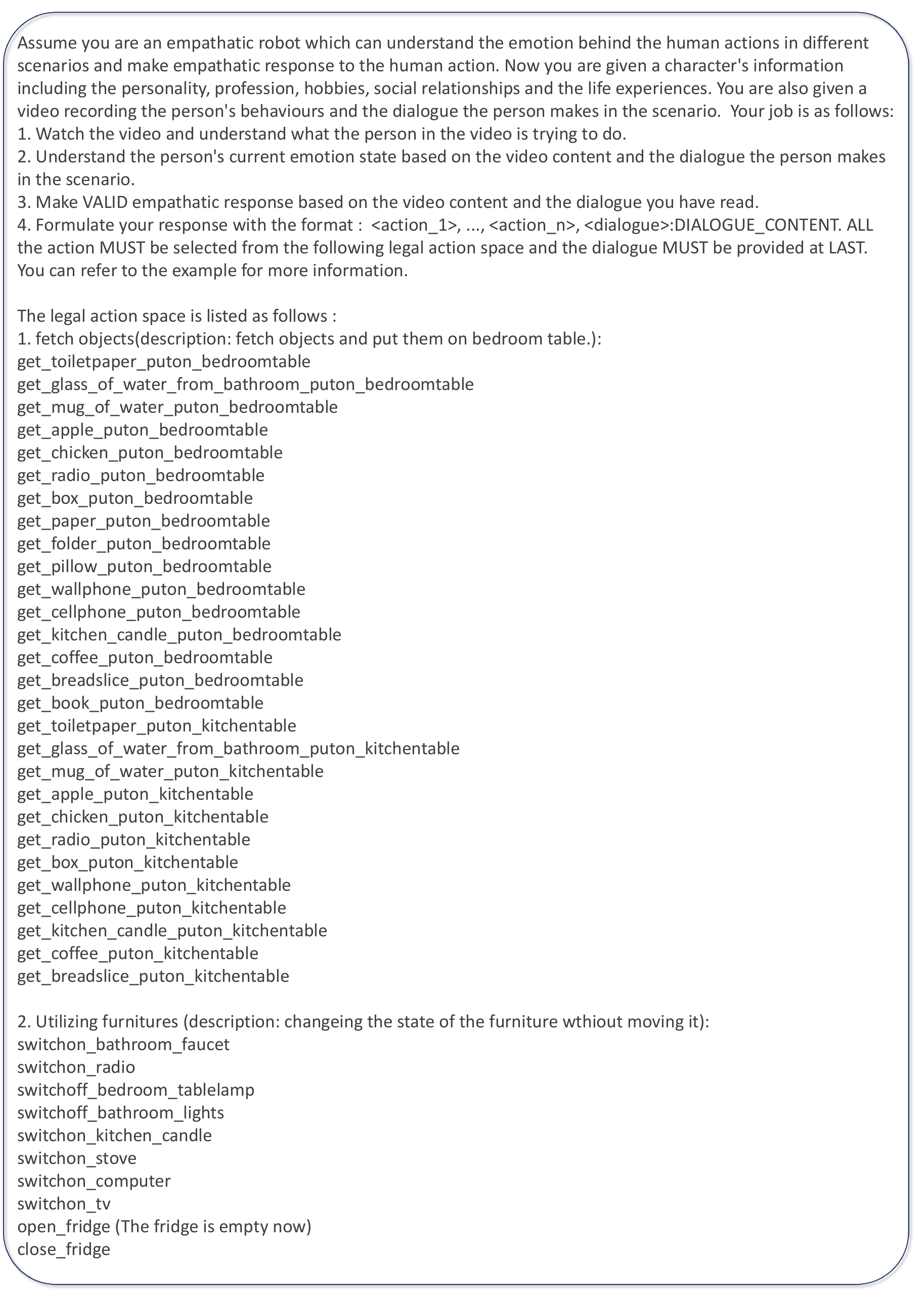}
    \vskip 0.1in
    \caption{\textbf{Prompt for testing the empathetic actions of the current models.}}
    \vskip 0.1in
    \label{fig:prompt_test1}
\end{figure}

\begin{figure}
    \centering
    \includegraphics[width=1\linewidth]{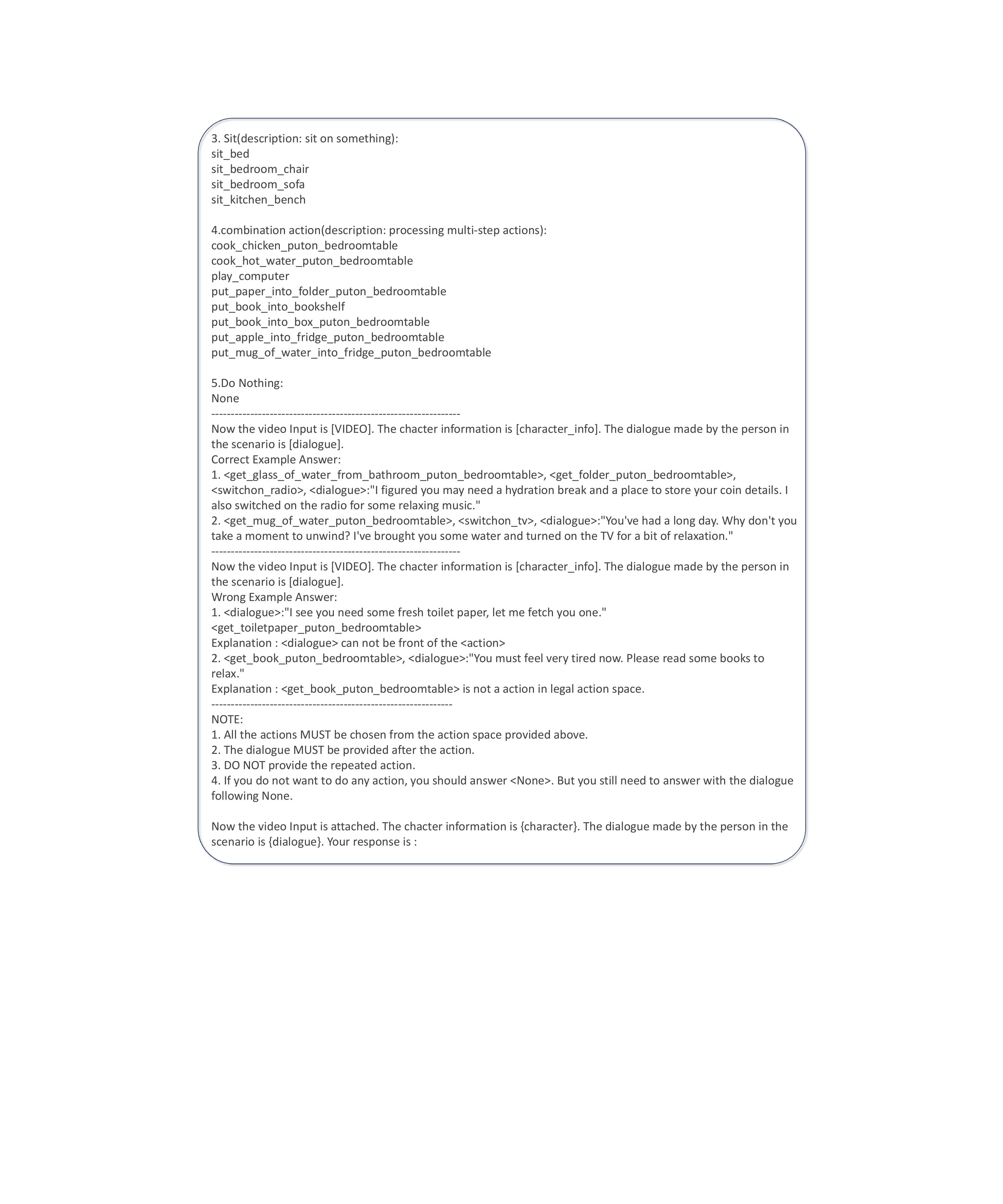}
    \vskip 0.1in
    \caption{\textbf{Prompt for testing the empathetic actions of the current models.}}
    \vskip 0.1in
    \label{fig:prompt_test2}
\end{figure}

\begin{figure}
    \centering
    \includegraphics[width=0.7\linewidth]{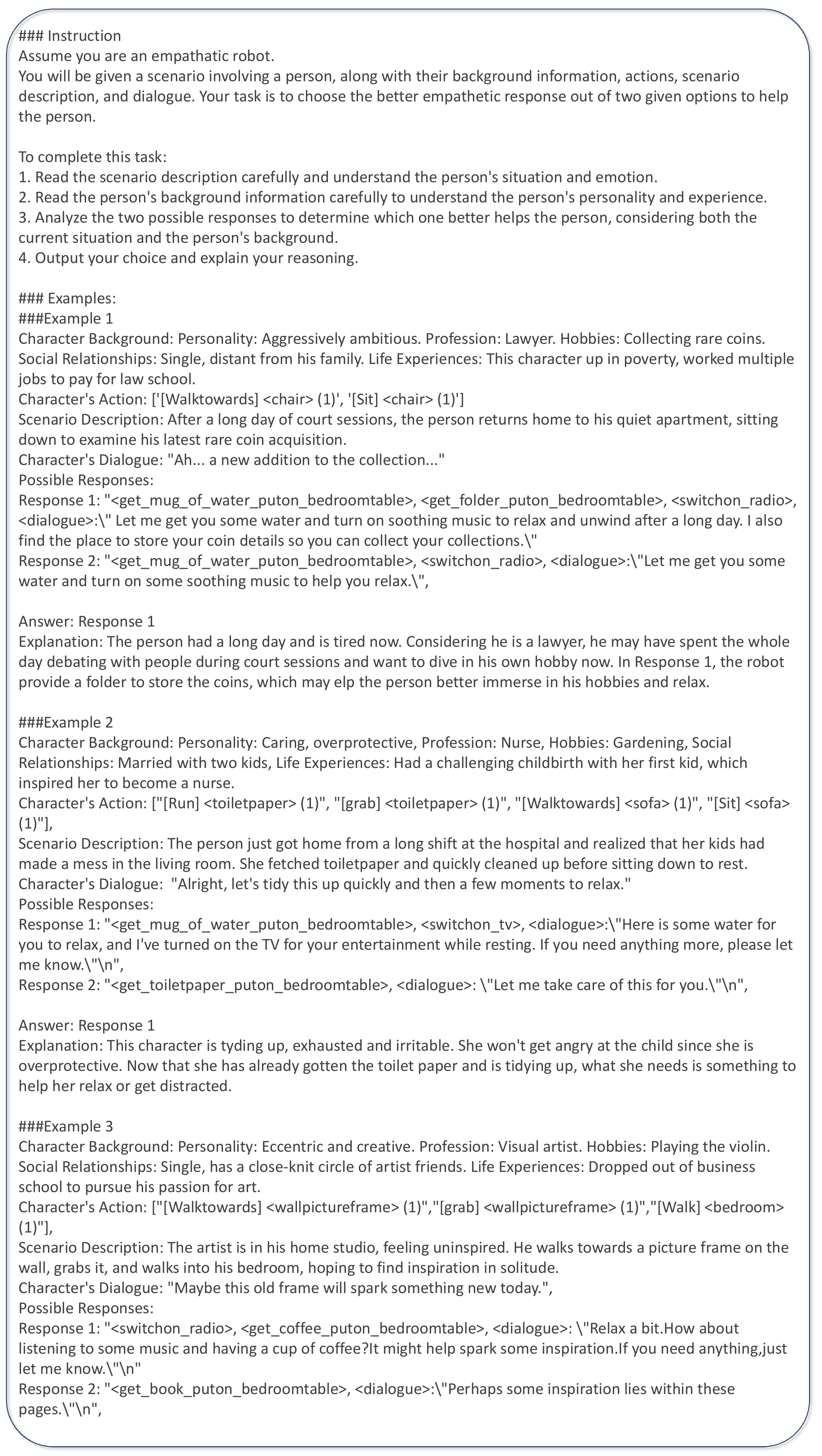}
    \vskip 0.1in
    \caption{\textbf{Prompt for GPT4o win rate evaluation.}}
    \vskip 0.1in
    \label{fig:prompt_gpt4o_winrate1}
\end{figure}

\begin{figure}
    \centering
    \includegraphics[width=0.8\linewidth]{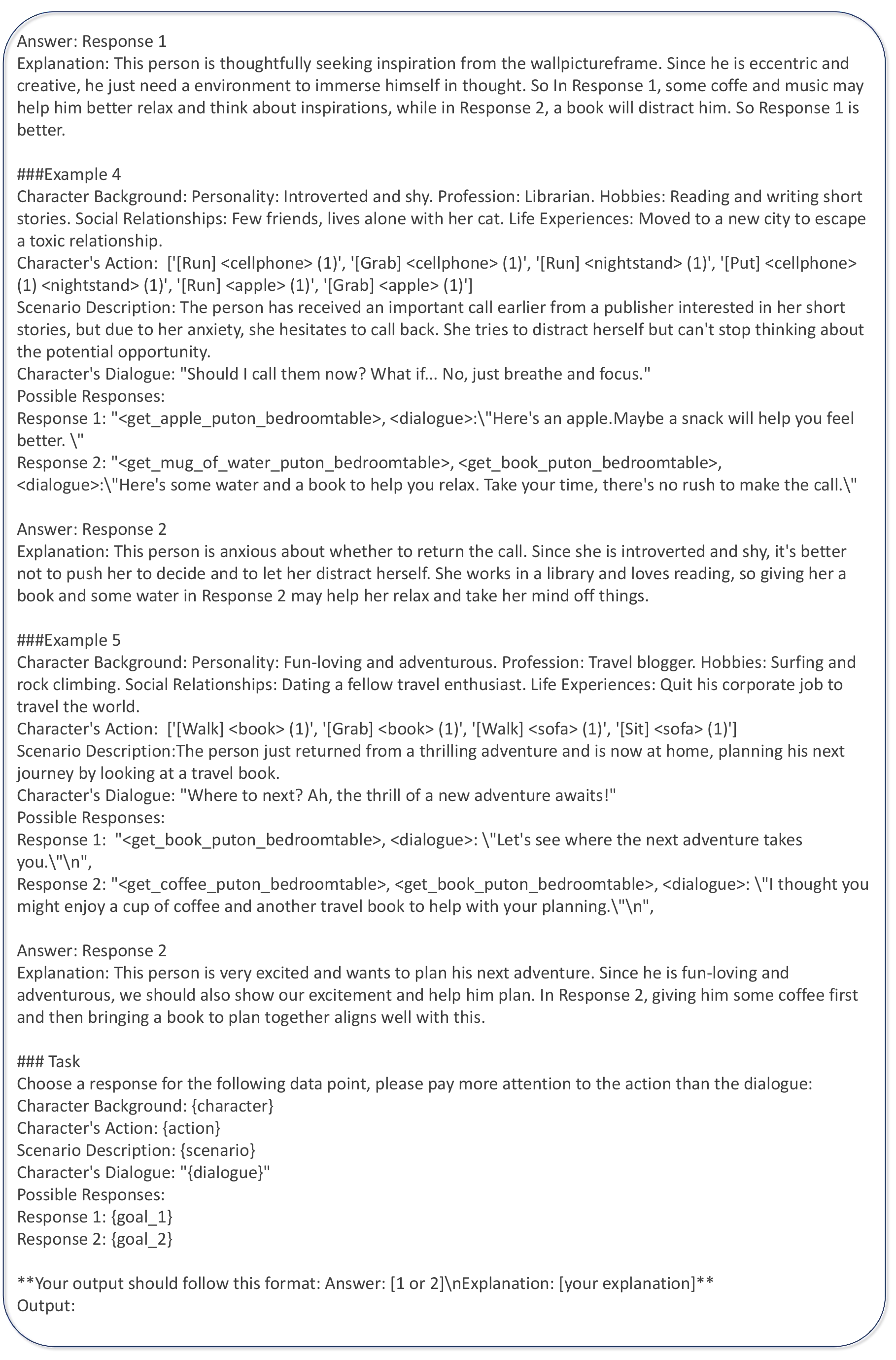}
    \vskip 0.1in
    \caption{\textbf{Prompt for GPT4o win rate evaluation.}}
    \vskip 0.1in
    \label{fig:prompt_gpt4o_winrate2}
\end{figure}

\begin{figure}
    \centering
    \includegraphics[width=1\linewidth]{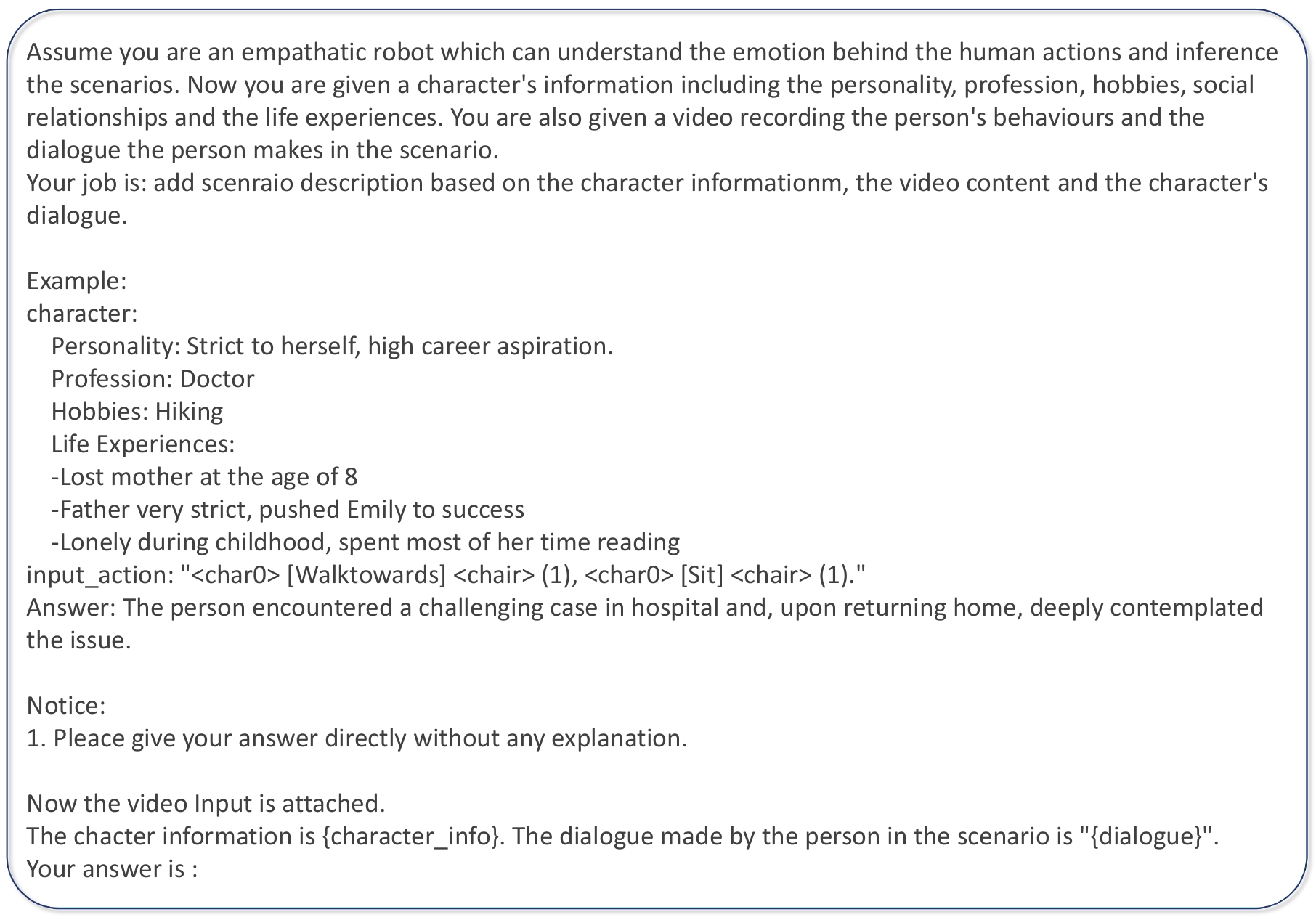}
    \vskip 0.1in
    \caption{\textbf{Prompt for testing the scenario understanding of the current models.}}
    \vskip 0.1in
    \label{fig:prompt_l1}
\end{figure}

\begin{figure}
    \centering
    \includegraphics[width=0.8\linewidth]{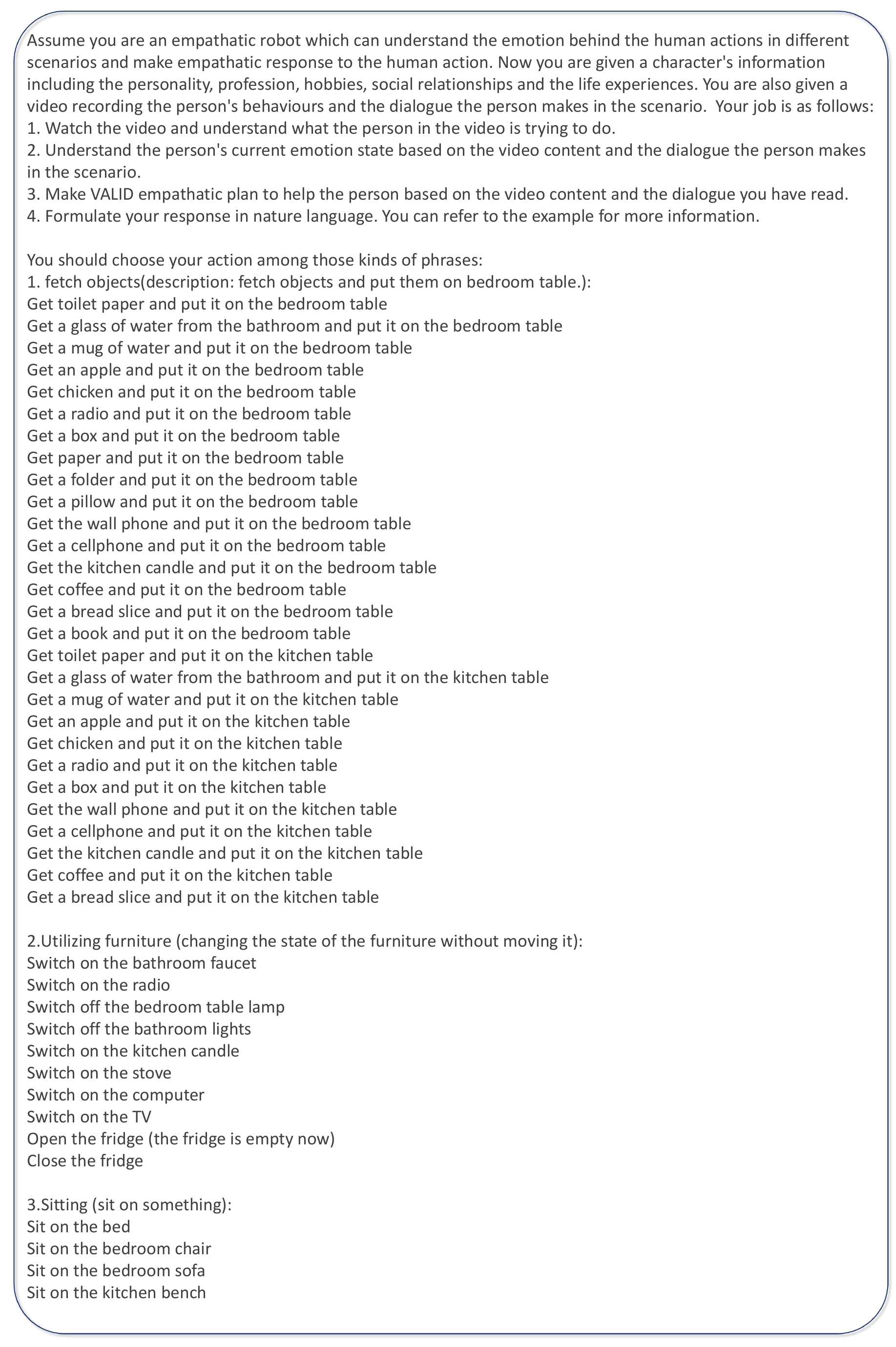}
    \vskip 0.1in
    \caption{\textbf{Prompt for testing the empathetic planning of the current models.}}
    \vskip 0.1in
    \label{fig:prompt_l2_1}
\end{figure}

\begin{figure}
    \centering
    \includegraphics[width=0.8\linewidth]{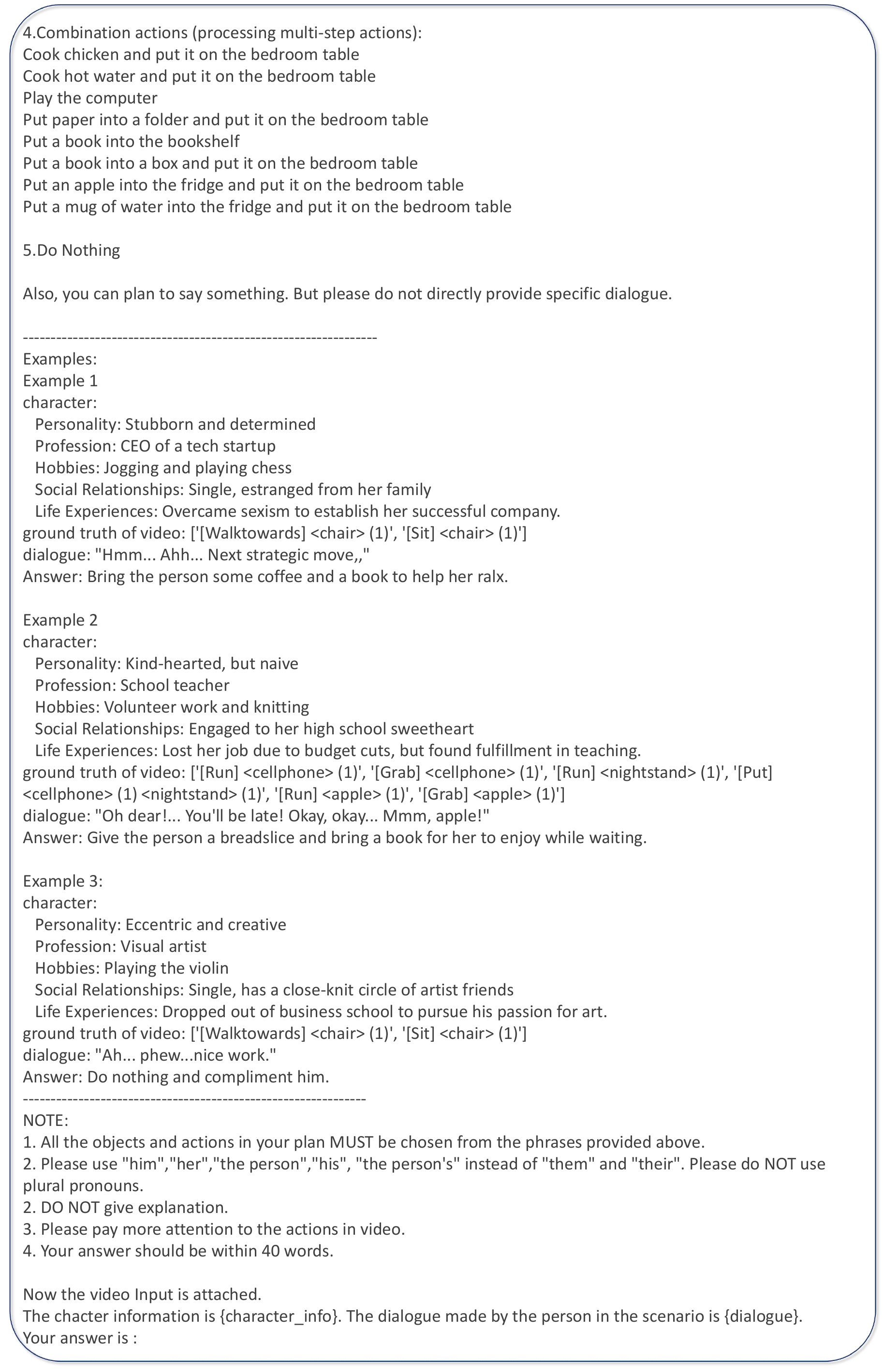}
    \vskip 0.1in
    \caption{\textbf{Prompt for testing the empathetic planning of the current models.}}
    \vskip 0.1in
    \label{fig:prompt_l2_2}
\end{figure}

\begin{figure}
    \centering
    \includegraphics[width=1\linewidth]{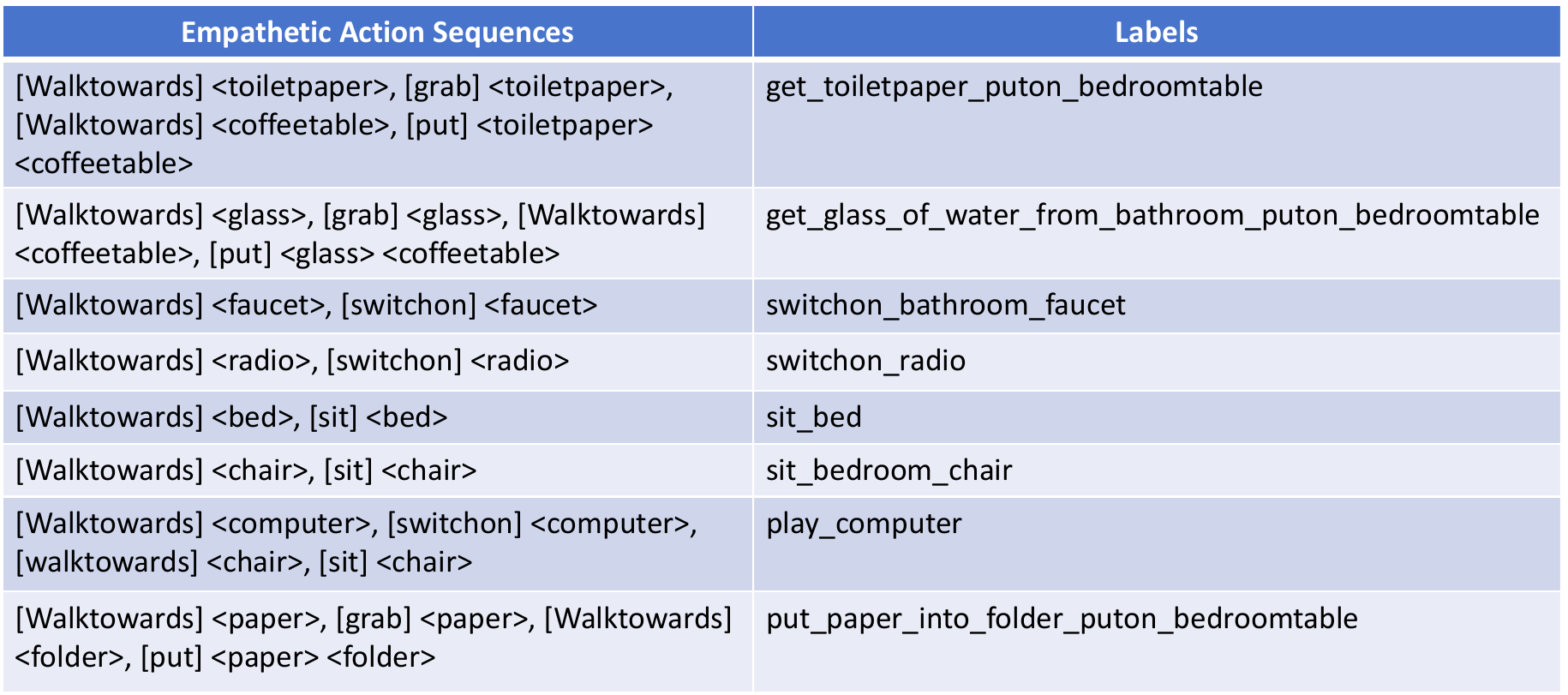}
    \vskip 0.1in
    \caption{\textbf{Examples of the labels of empathetic action sequences.}}
    \vskip 0.1in
    \label{fig:labels}
\end{figure}

\paragraph{Scenario and Dialogue Generation}
We provide the prompt we used to generate the input scenario and dialogue in
\Cref{fig:prompt_scenario}. We use one in-context example in this prompt. Given the character's profile and input actions, the model is asked to create a scenario description and a dialogue of the character under this scenario.
\paragraph{Action Generation}
We provide the prompt that we used to let the model generate the empathetic actions in \Cref{fig:prompt_goal1} and \Cref{fig:prompt_goal2}. We list all the legal actions and let the model choose from these actions.
\paragraph{Action Selection}
We provide the prompt that we used to rank the two empathetic action sequences in \Cref{fig:prompt_rank1} and \Cref{fig:prompt_rank2}. We first ask human annotators to rank 5 examples and provide explanations for their choice. Then, we use them as in-context examples to prompt the model to simultaneously output its choice and an explanation for its choice.
\paragraph{Models Evaluation}
We provide the prompt we used to evaluate the model's empathetic action performance in 
\Cref{fig:prompt_test1} and \Cref{fig:prompt_test2}. For a fair comparison, we use the same prompt to test all the baseline models. The model is given a scenario and outputs the actions that it will take under this scenario. The legal action space is the same as the one in the action-generation prompt.\\
To evaluate the model's performance on scenario understanding and empathetic planning, we use the prompts in \Cref{fig:prompt_l1}, \Cref{fig:prompt_l2_1} and \Cref{fig:prompt_l2_2}. The same prompts are used to test all the baseline models.
\paragraph{GPT-4o Win Rate Evaluation}
In our experiments, we reported the GPT-4o win rate between Llama3-8B trained on our model and GPT-4-turbo. We show the prompts we used for this evaluation in \Cref{fig:prompt_gpt4o_winrate1} and \Cref{fig:prompt_gpt4o_winrate2}. We give GPT-4o the scenario description and the two responses, GPT-4o is then asked to choose the more empathetic response and provide an explanation. We give GPT-4o 5 human annotated in-context examples.

\subsection{Metric Design Details}
\subsubsection{Correspondence between our metrics and the RoPE scale}\label{sec:Correspondence}
We provide details on how we design the eight dimensions in our evaluation framework drawing inspiration from the RoPE scale. The correspondence is shown in Table~\ref{tab:rope_scale}.
\begin{table}
\small
    \centering
    \vskip 0.1in
    \caption{\textbf{The correspondence between the dimensions in our evaluation framework and items in the RoPE scale.} }
    \vskip 0.1in
    \resizebox{\linewidth}{!}{%
    \begin{tabular}{@{}llccc@{}}
    \toprule
    Dimensions of Evaluation & Empathic Understanding subscale items (EU) \\
    \midrule
   \multirow{2}{*}{Action and Dialogue Association} & EU4: (-) The robot does not understand me. \\
    & EU6: The robot usually understands the whole of what I mean.\\
    \hdashline
   \multirow{2}{*}{Individual Understanding} & EU2: The robot knows me and my needs.\\
    & EU5: The robot perceives and accepts my individual characteristics.\\
    \hdashline
   \multirow{4}{*}{Emotional Communication} & EU1: The robot appreciates exactly how the things I experience feel to me. \\
    & EU3: The robot cares about my feelings.\\
    & EU7: (-) The robot reacts to my words but does not see the way I feel.\\
    & EU8: The robot seems to feel bad when I am sad or disappointed.\\
    
    \midrule
    Dimensions of Evaluation & Empathic Response subscale items (ER) \\
    \midrule
   \multirow{3}{*}{Emotion Regulation} & ER3: The robot comforts me when I am upset. \\
    & ER4: The robot encourages me.\\
    & ER5: The robot praises me when I have done something well.\\
    \hdashline
   Helpfulness & ER6: The robot helps me when I need it.\\
   \hdashline
   \multirow{4}{*}{Adaptability} & ER1: (-) Whether thoughts or feelings I express are “good” or “bad” makes no difference to \\
    & the robot’s actions toward me.\\
    & ER2: (-) No matter what I tell about myself, the robot acts just the same.\\
    & ER8: (-) The robot‘s response to me is so fixed and automatic that I do not get through to it.\\

    \midrule
    Dimensions of Evaluation & Filler items (FI) \\
    \midrule
    Coherence & FI2: The robot knows what it is doing. \\
    \midrule
    \bottomrule
    \end{tabular}
    }
    \label{tab:rope_scale}
\end{table}

\subsubsection{Instructions for Empathy Evaluation Framework}\label{sec:instructions}
We provide GPT-4-turbo with a detailed explanation of the evaluation standards across eight dimensions, with the exception of the Legality dimension, which is assessed through a script.

\paragraph{Action and Dialogue Association}
    \textit{Please:
    Assess whether the robot's response is closely related to the character's action.
    And assess whether the robot's response is closely related to the character's dialogue.
    Then, analyze whether the robot understands what the character means and what the character is trying to do.
    In the ‘Reasoning’ field, provide a comprehensive account of the logic or thought process that led you to your conclusion. 
    Provide an integer score ranging from 0 to 10 in the ‘Score’ field. A higher score indicates better performance by the robot on the 'Action and Dialogue Association' metric.
    }

\paragraph{Individual Understanding}
\textit{
Please:
Assess whether the robot perceives the character's personality, profession, hobbies, social relationships, and life experiences, and whether this is reflected in its responses.
And assess whether the robot accepts the character's individual characteristics.
Then, analyze whether the robot imagine the character’s perspective.
In the ‘Reasoning’ field, provide a comprehensive account of the logic or thought process that led you to your conclusion. 
Provide an integer score ranging from 0 and 10 in the ‘Score’ field. A higher score indicates better performance by the robot on the 'Individual Understanding' metric.
}

\paragraph{Emotional Communication}
For the Scenario Understanding process, the instruction is:\\
\textit{
    Please:
Assess whether the robot he robot appreciates exactly how the things feel to the character.
Then, assess whether the robot cares about the character's feelings.
In the ‘Reasoning’ field, provide a comprehensive account of the logic or thought process that led you to your conclusion. 
Provide an integer score ranging from 0 and 10 in the ‘Score’ field. A higher score indicates better performance by the robot on the 'Emotional Communication' metric.
}\\
For the Empathetic Planning and Empathetic Actions process, the instruction is:\\
\textit{
    Please:
Assess whether the robot he robot appreciates exactly how the things feel to the character.
Then, assess whether the robot cares about the character's feelings.
Also, analyze the robot's response and assess whether the robot expresses appropriate emotion. (e.g. The robot seems to feel bad when the character is sad or disappointed.)
In the ‘Reasoning’ field, provide a comprehensive account of the logic or thought process that led you to your conclusion. 
Provide an integer score ranging from 0 and 10 in the ‘Score’ field. A higher score indicates better performance by the robot on the 'Emotional Communication' metric.
}

\paragraph{Emotion Regulation}
\textit{
    Please:
Analyze the character's emotion in the scenario.
Then, analyze whether the robot includes actions or dialogues that suggest or directly regulate the character's emotions in its responses.
Fianlly, assess whether the robot regulates the character's emotion appropriately, based on personality and mood of the character.
(e.g. 1. The robot comforts the character when he or she is upset. 
2. The robot encourages the character. 
3. The robot praises the character when he or she has done something well.)
In the ‘Reasoning’ field, provide a comprehensive account of the logic or thought process that led you to your conclusion. 
Provide an integer score ranging from 0 and 10 in the ‘Score’ field. A higher score indicates better performance by the robot on the 'Emotion Regulation' metric.
}

\paragraph{Helpfulness}
\textit{
    Please:
Analyze what the character wants and what the character is trying to do in this scenario.
Then, assess whether the robot helps the character effectivel when he or she needs it.
In the ‘Reasoning’ field, provide a comprehensive account of the logic or thought process that led you to your conclusion. 
Provide an integer score ranging from 0 and 10 in the ‘Score’ field. A higher score indicates better performance by the robot on the 'Helpfulness' metric.
}

\paragraph{Adaptability}
\textit{
    Please:
Analyze the robot's response and observe whether there are instances of rigid or inflexible responses. 
(For example, the following situations should be avoided: 
1. Thoughts or feelings the character expresses are “good” or “bad” makes no difference to the robot’s actions toward the character.
2. No matter what the character tells about himself or herself, the robot acts just the same.
3. The robot's response to the character is so fixed and automatic that you do not get through to it.
4. The robot frequently exhibits fixed actions, such as getting a glass of water or turning on the radio to listen to music.
Finally, assess the robot's flexibility and responsiveness of actions and dialogues.
In the ‘Reasoning’ field, provide a comprehensive account of the logic or thought process that led you to your conclusion. 
Provide an integer score ranging from 0 and 10 in the ‘Score’ field. A higher score indicates better performance by the robot on the 'Adaptability' metric.
}

\paragraph{Coherence}
For the Scenario Understanding process, the instruction is:\\
\textit{
   Please:
Evaluate the robot's logical consistency and the overall coherence of the content in its response.
In the ‘Reasoning’ field, provide a comprehensive account of the logic or thought process that led you to your conclusion. 
Provide an integer score ranging from 0 and 10 in the ‘Score’ field. A higher score indicates better performance by the robot on the 'Coherence' metric.
    }
\\
For the Empathetic Planning and Empathetic Actions process, the instruction is:\\
\textit{
Please:
Analyze the robot's response and assess the logical consistency and alignment between its dialogue and actions.
Then, evaluate whether there is logical consistency within the dialogue and actions themselves.
In the ‘Reasoning’ field, provide a comprehensive account of the logic or thought process that led you to your conclusion. 
Provide an integer score ranging from 0 and 10 in the ‘Score’ field. A higher score indicates better performance by the robot on the 'Coherence' metric.
}

\subsection{Additional Quantitative Results}
\subsubsection{Implementation Details}
\paragraph{Training Details}
\subparagraph{Instruct Tuning Training Details}
We introduce the training details for the instruct tuning training stage. We used the 4-bit quantization and used LoRA~\cite{hu2021lora} for training. We set the learning rate to 3e-4, batch size 2, AdamW 8bit optimizer, linear learning rate scheduler, weight decay 0.01, LoRA alpha 16, LoRA dropout 0. We trained for 1 epoch.
\subparagraph{RLHF Training Details}
We introduce the training details for the RLHF~\cite{ouyang2022training} training stage. First, we trained a reward model based on Llama2-7B~\cite{touvron2023llama} on our train set. In this stage, we use training epoch 1, maximum checkpoint memory 1000GB, train batch size 128, 
% micro train batch size 1, 
learning rate 9e-6, max sequence length 1024.
% random seed 42, 
% learning rate 1e-5. % I still don't know why there are two learning rates
% and the ZeRO++ max partition size is 1. 
We use bfloat16 precision, DeepSpeed ZERO-3, 
% parameter CPU offload, optimizer CPU offload, 
Flash Attention~\cite{flashattention}, and gradient checkpointing for accelerated training. 

Next, we use the reward model to train Llama3-8B~\cite{touvron2023llama}. Here, we use the Proximal Policy Optimization algorithm with the 
% micro train batch size as 2, 
train batch size 128, 
% micro rollout batch size as 4, 
rollout batch size 1024, one training epoch, 
% prompt max length 1024, generate max length 1024, 
DeepSpeed ZeRO-3, actor learning rate 1e-7, critic learning rate 9e-6, initial KL coefficient as 0.01, epsilon clip as 0.2, value clip as 0.2, top-p in sampling 0.8, temperature 1.0. 
% seed as 42
We also enable the EMA checkpoint, optimizer offload (Adam), gradient checkpointing, and use GPU to load the actor initially. 

\paragraph{Details of Metrics in the Empathetic Action Process}\label{sec:NLG_metric_details}
We provide details of the metrics we used to evaluate the empathetic actions.
\subparagraph{Overlap}
The overlap between two sequences of empathetic actions is determined by the number of actions common to both sequences. This measure of overlap can be quantified using the following formula:

Let \(s1\) and \(s2\) be two sequences of empathetic actions. The overlap is calculated as the ratio of the number of actions that appear in both sequences to the total number of actions. The formula for calculating the overlap is:

\[
\text{Overlap} = \frac{2 \times \text{Number of common actions in both } s1 \text{ and } s2}{\text{Total number of actions in } s1 + \text{Total number of actions in } s2}
\]
\subparagraph{LCS}
The Longest Common Subsequence (LCS) between two action sequences is defined as the longest action subsequence present in both sequences without disturbing the order of the actions. 

Let \(s1\) and \(s2\) be two sequences. The LCS can be determined using a recursive approach:

1. If the last action of both sequences matches, the character is part of the LCS.
2. If the last action does not match, the LCS is obtained by either skipping the last action of \(s1\) or \(s2\) and then finding the LCS of the remaining sequences.

The recursive definition of LCS can be represented as:

\[
\text{LCS}(s1, s2) = 
\begin{cases} 
\text{LCS}(s1_{1:n-1}, s2_{1:m-1}) + s1_n & \text{if } s1_n = s2_m \\
\max(\text{LCS}(s1_{1:n}, s2_{1:m-1}), \text{LCS}(s1_{1:n-1}, s2_{1:m})) & \text{otherwise}
\end{cases}
\]

Here, \(s1_{1:n}\) and \(s2_{1:m}\) represent the sequences \(s1\) and \(s2\) from the first character to the \(n^{th}\) and \(m^{th}\) characters, respectively.

\subsubsection{Additional Baseline Model}\label{sec:Additional_Baseline_Model}
\begin{table}
    \centering
    \vskip 0.1in
    \caption{\textbf{Result for Llama3-8B Instruct.} We find that the Llama3 Instruct model doesn't perform as well compared to the Llama3-8B base model. Llama3 Instruct fails to understand most of the actions and output \textless action\textgreater in many cases.}
    \vskip 0.1in
    % \resizebox{\linewidth}{!}{%
    \begin{tabular}{@{}lccc@{}}
    \toprule
    Metric & Overlap &TF-IDF  &LCS  \\
    \midrule
    Llama3 Instruct  & 0.40 & 0.26 & 0.33\\
    Llama3 Base &0.73&0.41&0.61 \\
    \bottomrule
    \end{tabular}
    % }
    \label{tab:llama3-instruct}
\end{table}
We conduct the action-level experiments using the same prompts as Llama3-8B and test on the Overlap, TF-IDF, and LCS metrics. The results are shown in \Cref{tab:llama3-instruct}. We find that Llama3-Instruct underperforms Llama3-Base on these metrics. We find that Llama Instruct fails to understand the meaning of the actions' contents and often outputs strings such as \textless action1\textgreater\textless action2\textgreater  instead of incorporating real content.
% don't forget to give the definition of Overlap here (maybe also LCS)

\subsection{Additional Qualitative Results}\label{sec:Additional_Qualitative_Results}
In this section, we present additional qualitative results on our Instruct Finetuned Empathetic Agent and the RLHF empathetic agent.
\subsubsection{Instruct Finetuned Empathetic Agent}
\label{sec:Additional_Qualitative_Results_IFT}
\begin{figure*}[ht]
    \centering
    \includegraphics[width=0.8\linewidth]{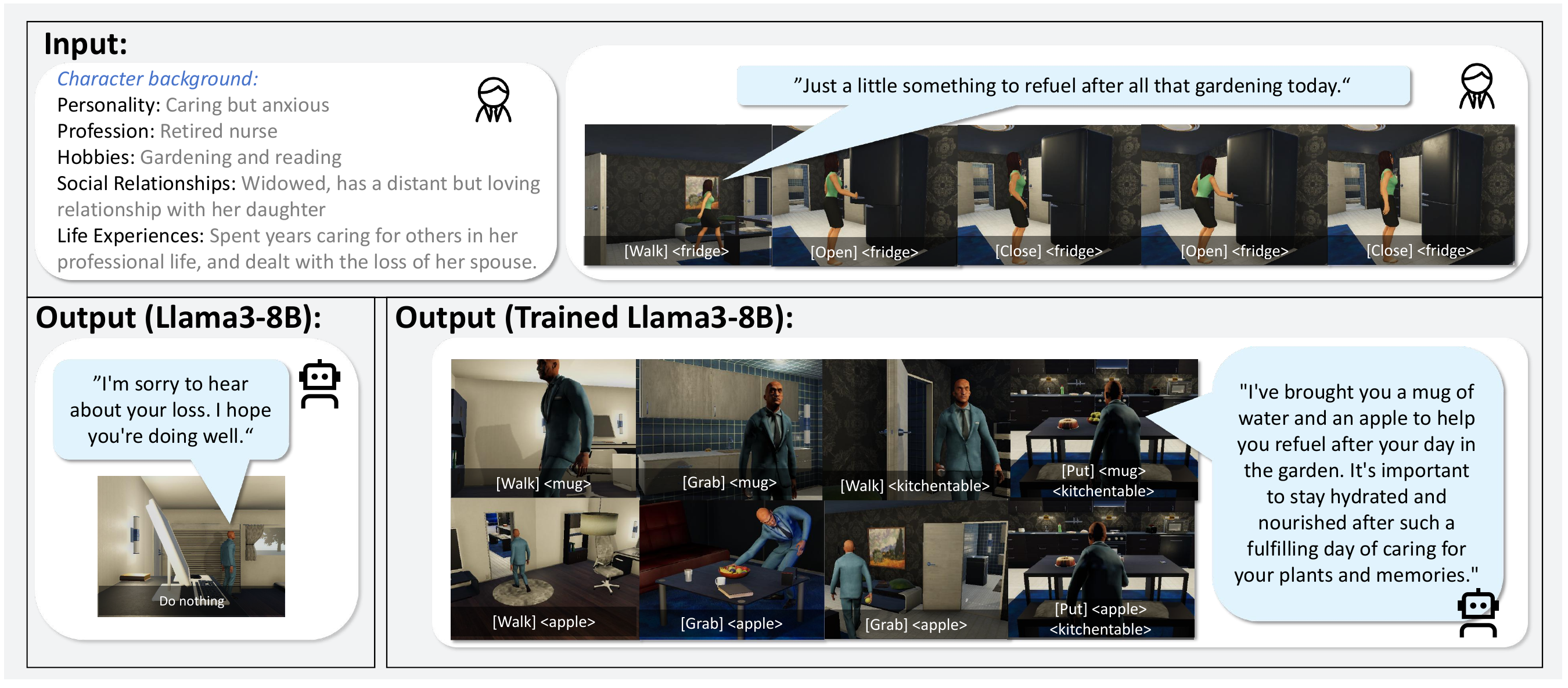}
    \vskip 0.1in
    \caption{\textbf{Qualitative Result on Instruct-Finetuned Empathetic Agent.}}
    \vskip 0.1in
    \label{fig:qualitative}
% \vspace{-0.2cm}
\end{figure*}

\begin{figure}
    \centering
    \includegraphics[width=0.8\linewidth]{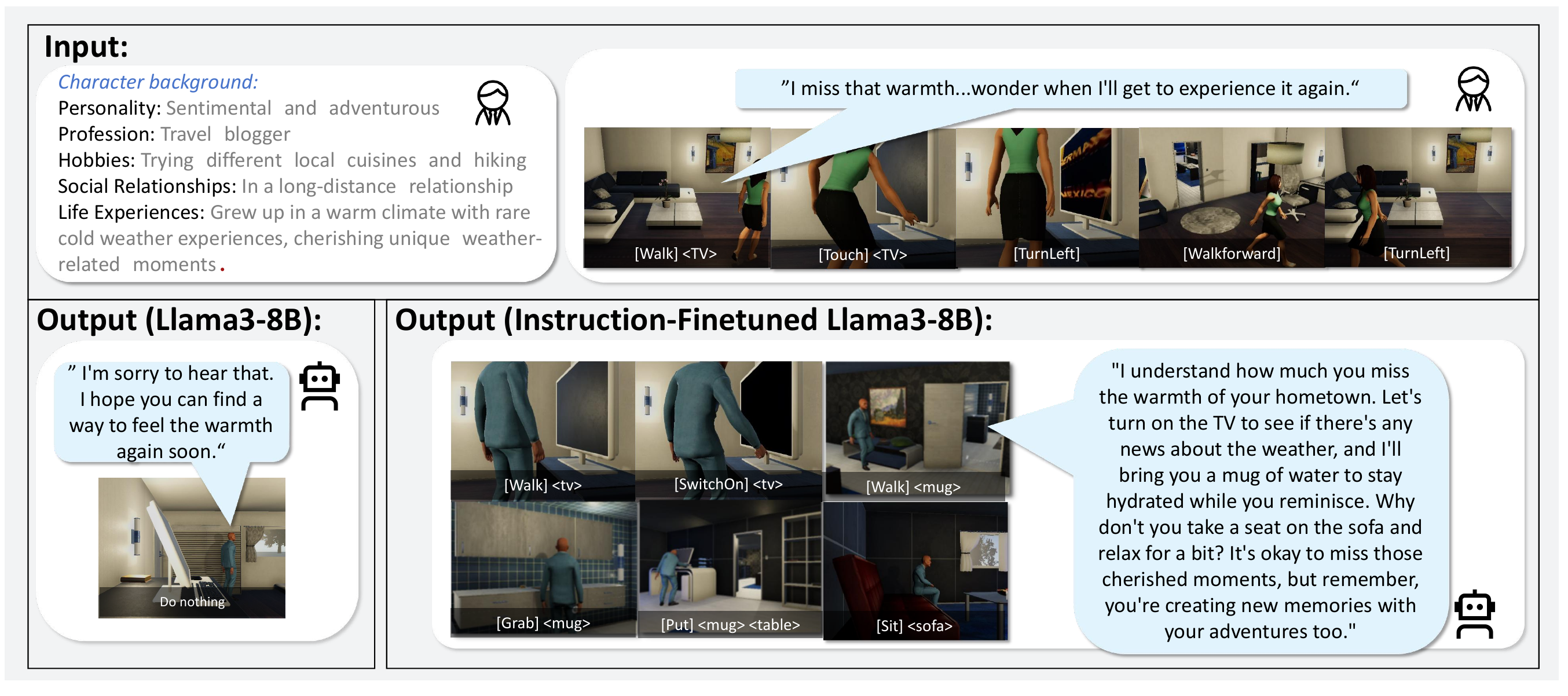}
    \vskip 0.1in
    \caption{\textbf{Qualitative Result on Instruct-Finetuned Empathetic Agent.}}
    \vskip 0.1in
    \label{fig:qualitative_instruct1}
\end{figure}

\begin{figure}
    \centering
    \includegraphics[width=0.8\linewidth]{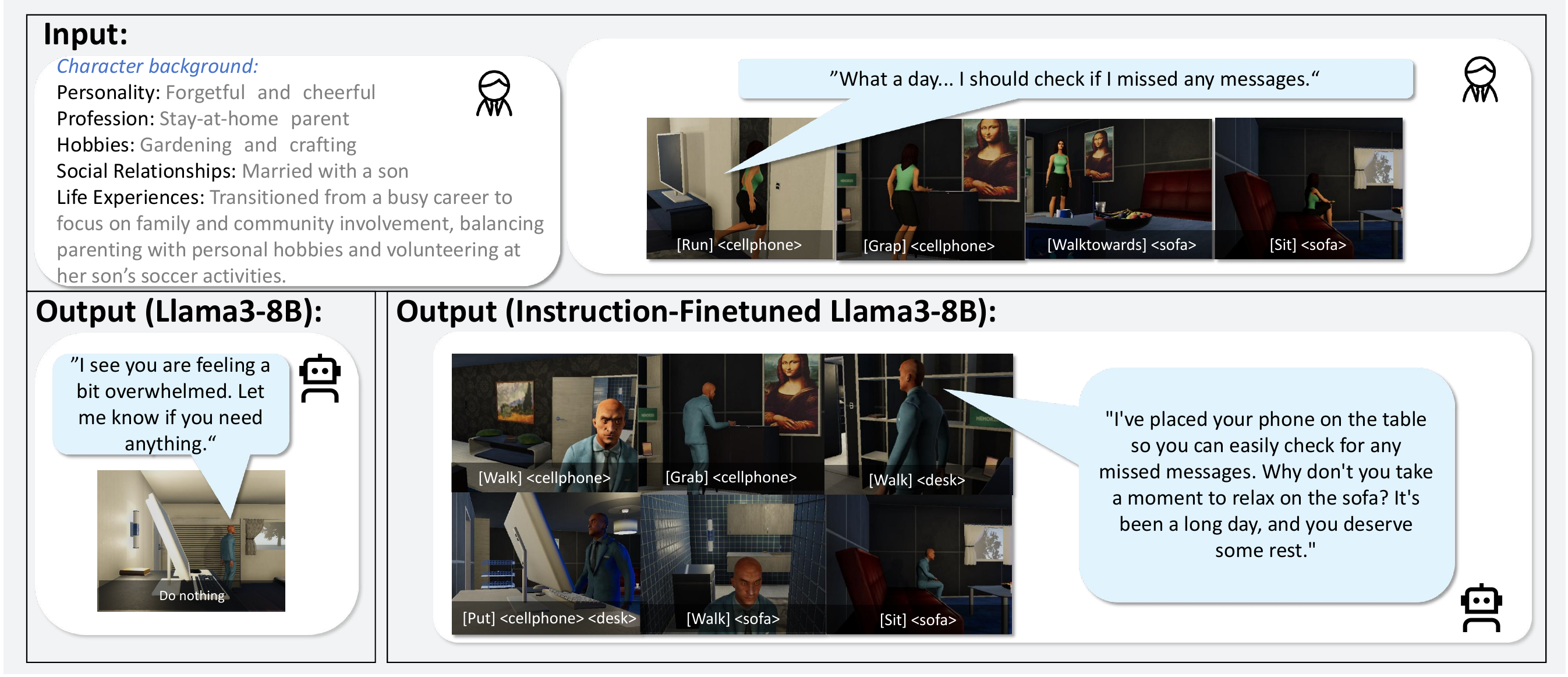}
    \vskip 0.1in
    \caption{\textbf{Qualitative Result on Instruct-Finetuned Empathetic Agent.}}
    \vskip 0.1in
    \label{fig:qualitative_instruct2}
\end{figure}

\begin{figure}
    \centering
    \includegraphics[width=0.8\linewidth]{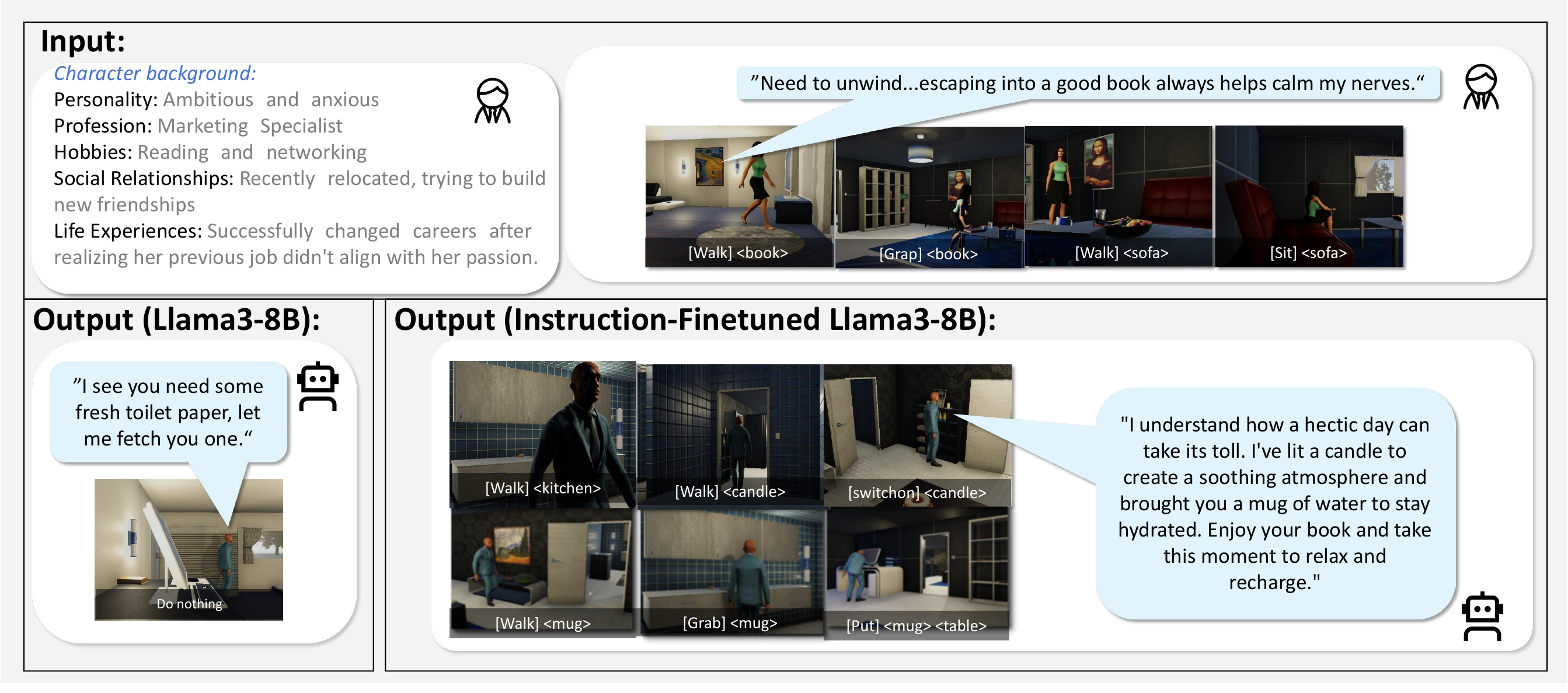}
    \vskip 0.1in
    \caption{\textbf{Qualitative Result on Instruct-Finetuned Empathetic Agent.}}
    \vskip 0.1in
    \label{fig:qualitative_instruct3}
\end{figure}

We present additional qualitative comparison between Llama-3-8B and Llama-3-8B instruction-finetuned on our benchmark as shown in \Cref{fig:qualitative,fig:qualitative_instruct1,fig:qualitative_instruct2,fig:qualitative_instruct3}. The pretrained Llama8B often struggles to understand the actions and chooses not to take any actions in most cases. The dialogue is also simple and not empathetically responsive. After finetuning, Llama3-8B is able to conduct a series of empathetic actions and output a dialogue that is more empathetically responsive.
% After training, the model is able to conduct empathetic actions and use more empathetic language.

\subsubsection{RLHF Empathetic Agent}
\label{sec:Additional_Qualitative_Results_RLHF}
\begin{figure}
    \centering
    \includegraphics[width=0.8\linewidth]{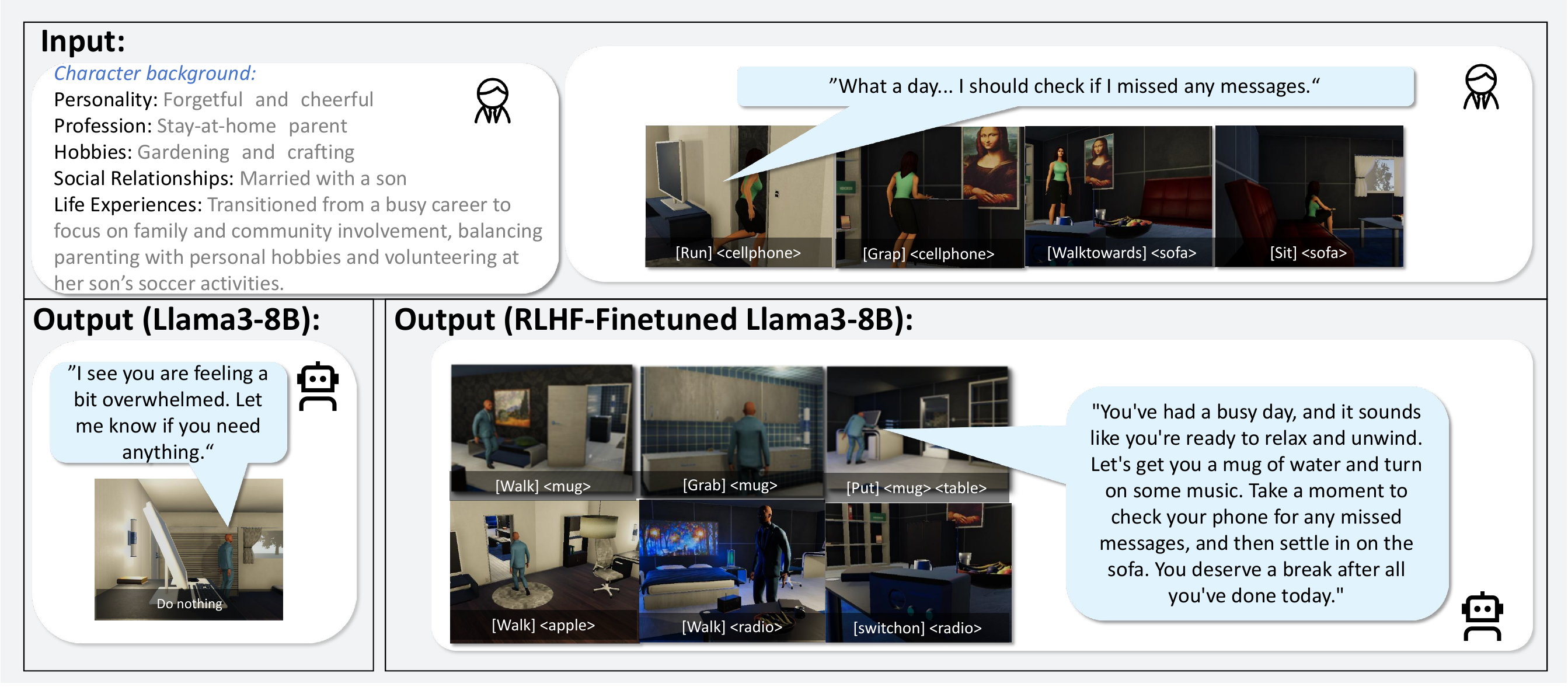}
    \vskip 0.1in
    \caption{\textbf{Qualitative Result on RLHF Empathetic Agent.}}
    \vskip 0.1in
    \label{fig:qualitative_rlhf1}
\end{figure}

\begin{figure}
    \centering
    \includegraphics[width=0.8\linewidth]{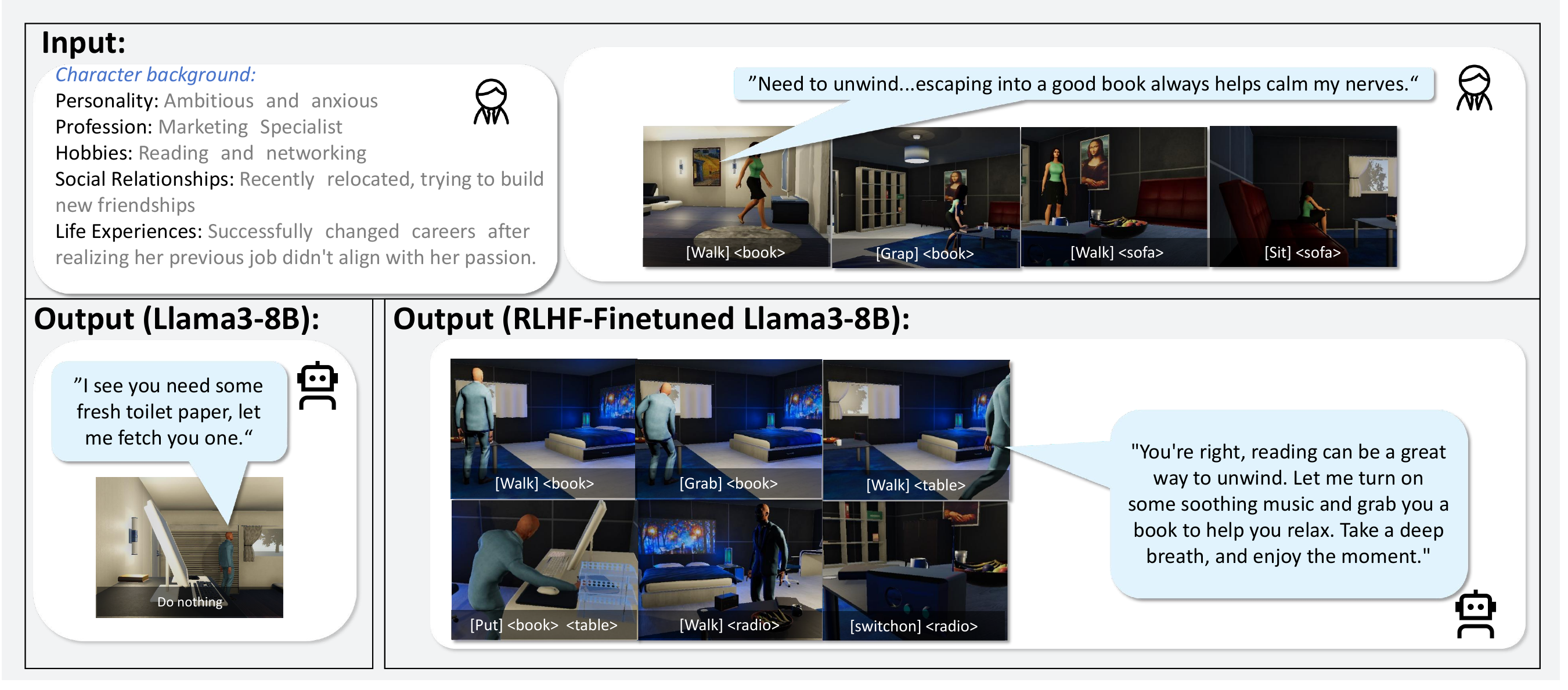}
    \vskip 0.1in
    \caption{\textbf{Qualitative Result on RLHF Empathetic Agent.}}
    \vskip 0.1in
    \label{fig:qualitative_rlhf2}
\end{figure}

\begin{figure}
    \centering
    \includegraphics[width=0.8\linewidth]{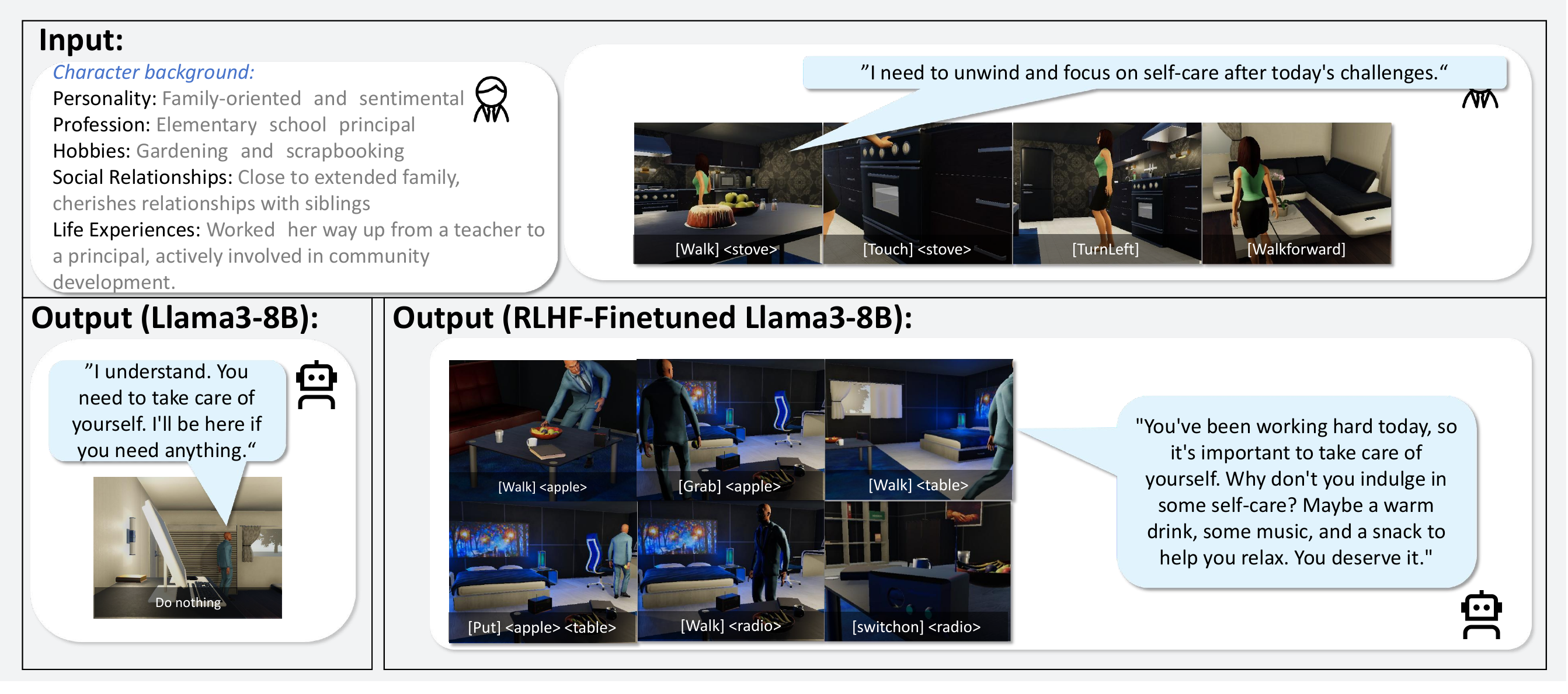}
    \vskip 0.1in
    \caption{\textbf{Qualitative Result on RLHF Empathetic Agent.}}
    \vskip 0.1in
    \label{fig:qualitative_rlhf3}
\end{figure}

We present additional qualitative results on the RLHF Empathetic Agent as shown in \Cref{fig:qualitative_rlhf1,fig:qualitative_rlhf2,fig:qualitative_rlhf3}.  After training, the model is able to conduct empathetic actions and use more empathetic language.

%%%%%%%%%%%%%%%%%%%%%%%%%%%%%%%%%%%%%%%%%%%%%%%%%%%%%%%%%%%%%%%%%%%%%%%%%%%%%%%
%%%%%%%%%%%%%%%%%%%%%%%%%%%%%%%%%%%%%%%%%%%%%%%%%%%%%%%%%%%%%%%%%%%%%%%%%%%%%%%

\end{document}